\documentclass[sigconf]{acmart}
\AtBeginDocument{%
  \providecommand\BibTeX{{%
    \normalfont B\kern-0.5em{\scshape i\kern-0.25em b}\kern-0.8em\TeX}}}

\copyrightyear{2025}
\acmYear{2025}
\setcopyright{cc}
\setcctype{by}
\acmConference[CHI '25]{CHI Conference on Human Factors in Computing Systems}{April 26-May 1, 2025}{Yokohama, Japan}
\acmBooktitle{CHI Conference on Human Factors in Computing Systems (CHI '25), April 26-May 1, 2025, Yokohama, Japan}
\acmDOI{10.1145/3706598.3714002}
\acmISBN{979-8-4007-1394-1/25/04}

\usepackage{array}
\newcolumntype{H}{>{\setbox0=\hbox\bgroup}c<{\egroup}@{}}

\usepackage{algorithm,algorithmic}

\usepackage{multirow}
\usepackage{enumitem}

\usepackage[]{color-edits}
\addauthor{vc}{blue}

\begin{document}

\title{Need Help? Designing Proactive AI Assistants for Programming}
\author{Valerie Chen}
\affiliation{%
  \institution{Carnegie Mellon University}
  \city{Pittsburgh}
  \state{PA}
  \country{USA}
}
\email{valeriechen@cmu.edu}

\author{Alan Zhu}
\affiliation{%
  \institution{Carnegie Mellon University}
  \city{Pittsburgh}
  \state{PA}
  \country{USA}
}
\email{alanzhuyixuan@gmail.com}

\author{Sebastian Zhao}
\affiliation{%
  \institution{UC Berkeley}
  \city{Berkeley}
  \state{CA}
  \country{USA}
}
\email{sebbyzhao@berkeley.edu}

\author{Hussein Mozannar}
\affiliation{%
 \institution{Microsoft Research}
 \city{Redmond}
 \state{WA}
 \country{USA}}
 \email{hmozannar@microsoft.com}

\author{David Sontag}
\affiliation{%
  \institution{Massachusetts Institute of Technology}
  \city{Cambridge}
  \state{MA}
  \country{USA}}
\email{dsontag@mit.edu}

\author{Ameet Talwalkar}
\affiliation{%
  \institution{Carnegie Mellon University}
  \city{Pittsburgh}
  \state{PA}
  \country{USA}
}
\email{talwalkar@cmu.edu}

\renewcommand{\shortauthors}{Chen et al.}

\begin{abstract}
While current chat-based AI assistants primarily operate reactively, responding only when prompted by users, there is significant potential for these systems to proactively assist in tasks without explicit invocation, enabling a mixed-initiative interaction. This work explores the design and implementation of proactive AI assistants powered by large language models.
We first outline the key design considerations for building effective proactive assistants. As a case study, we propose a proactive chat-based programming assistant that automatically provides suggestions and facilitates their integration into the programmer's code. The programming context provides a shared workspace enabling the assistant to offer more relevant suggestions.
We conducted a randomized experimental study examining the impact of various design elements of the proactive assistant on programmer productivity and user experience. Our findings reveal significant benefits of incorporating proactive chat assistants into coding environments, while also uncovering important nuances that influence their usage and effectiveness.

\end{abstract}

\ccsdesc[500]{Human-centered computing~User studies}
\ccsdesc[500]{Software and its engineering~Collaboration in software development}

\keywords{AI-assisted Programming, Proactivity, Mixed-Initiative Interaction}

\begin{teaserfigure}
\centering
\includegraphics[width=\textwidth]{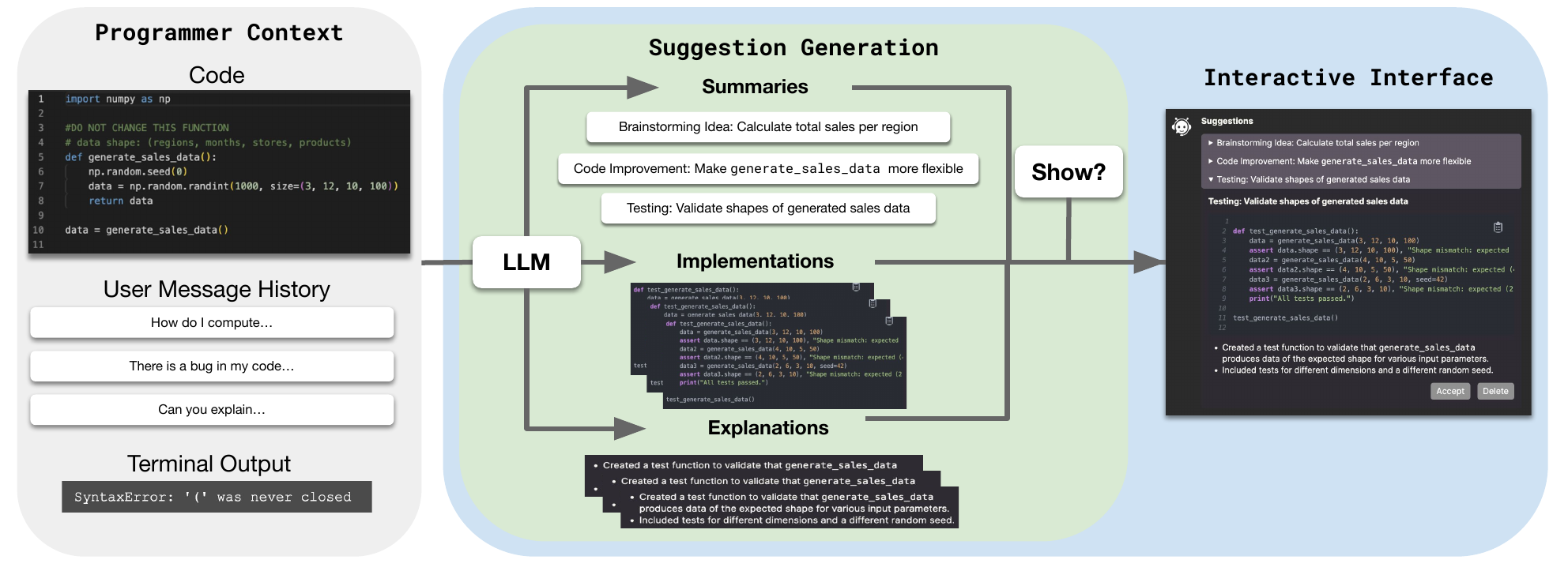}
\caption{\textbf{The implementation of a proactive chat assistant for programming.} We introduce a proactive assistant that takes in the programmer's context, which includes the current code, user message history, and optionally terminal output, generates a set of suggestions, which include a summary, implementation, and explanation of implementation, and then determines whether it is timely to show the user in the interactive interface.}
\Description{\textbf{The implementation of a proactive chat assistant for programming.} The proactive assistant takes in the programmer's context, which includes the current code, user message history, and optionally terminal output, generates a set of suggestions, which include a summary, implementation, and explanation of implementation, and then determines whether it is timely to show the user in the interactive interface.}
\label{fig:system_overview}
\end{teaserfigure}


\maketitle

\section{Introduction}\label{sec:intro}

Chat-based AI assistants such as ChatGPT~\cite{chatgpt} or Claude~\cite{claude}, enable people to accomplish some of their tasks via prompting: the user crafts a message with sufficient context about their task and then receives an AI response, they continue iterating this process until they're satisfied with the result.  Common use cases of chat-based AI assistance include writing tasks~\cite{lee2022coauthor, lee2024design}, idea generation \cite{shaer2024ai}, programming tasks \cite{xiao2023devgpt,mozannar2022reading}, and web navigation among other tasks \cite{lin2022inferring,yao2024tau}. Interacting with these assistants requires two points of effort from the human: first porting over the workspace context, e.g., copying the document or code file, and second describing the task in natural language. Assuming the assistant had access to the working context of the user, a natural question is whether it could \emph{infer the task they want to solve.} 
If the assistant can reliably provide suggestions relevant to the user's task, then it might be advantageous for it to generate suggestions automatically.
The assistant would now be \emph{proactive} and result in a mixed-initiative interaction \cite{horvitz1999principles}. In this paper, we attempt to design effective proactive chat-based assistants powered by large language models (LLMs) for programming.

Virtual assistants have been explored in many commercial products including Apple Siri and Amazon Alexa~\cite{luger2016like}; in fact, one of the most well-known and first proactive virtual assistants is Microsoft's Office Assistant, with the English version commonly referred to as Clippy. The Office Assistant would proactively offer its assistance based on inference of the user actions in their Word document.
While the user experience with Clippy was less than ideal~\cite{baym2019intelligent}, more recent examples of proactive assistants in the form of autocompletion for text-based tasks have been more successful.
For instance, GitHub Copilot \cite{copilot} provides in-line code suggestions with LLMs. Copilot has been well received by programmers and randomized studies have shown increases in programmer productivity \cite{vaithilingam2022expectation,peng2023impact}.
However, autocomplete-based proactive assistance only considers local suggestions based on the user's cursor position. In contrast, a chat-based proactive assistant could suggest modifying the entire code or pointing out bugs in earlier parts of the code, in addition to being able to complete the code. Moreover, it is clear that the particular design details of a proactive assistant are crucial to its effectiveness.

We begin by outlining design considerations for designing chat-based proactive assistants with LLMs to address challenges of proactivity~\cite{chaves2021should} by building on previous guidelines for human-AI interaction and mixed-initiative interaction~\cite{horvitz1999principles,amershi2019guidelines,lee2024design}. 
Key considerations include supporting efficient evaluation and utilization of proactive suggestions as well as timing of suggestions based on context. 
The domain of programming presents a good opportunity to study proactivity given the success of LLM-based assistants for programming and a well-defined work context, which consists of the code file(s). This enables us to overcome the challenge of observing the human's workspace by integrating the assistant into the programmer's integrated development environment (IDE). 
Based on these design considerations, we build a proactive assistant for programming and integrate it into a chat interface inside a code editor, as shown in Figure~\ref{fig:system_overview}. The chat interface operates like a standard interface where the user can type messages and receive responses; however, the assistant can now send proactive messages without a corresponding user request. 
The timing of the proactive suggestions depends on the human's recent activities in the editor and their interactions with the assistant responses.
The proactive messages are distinguished from regular assistant messages, with the former only displaying a high-level description of the message unless the user chooses to expand the message. 
Finally, each assistant message can be integrated into the code via a ``preview'' button that highlights the changes the message will make to the code.

We perform a controlled user study where we evaluate programmer usage of our proactive assistant against a baseline ChatGPT-like assistant.
We use the web-based code editor RealHumanEval~\cite{mozannar2024realhumaneval} to experiment as it allows us to easily conduct studies.
In the study, we include multiple conditions varying different aspects of the proactive assistant, including the timing of the suggestions and the ability to integrate suggestions into the programmer's code, to understand the effects of changing different design decisions. 
Our study demonstrates the benefits of proactivity in chat assistants for code as proactive assistants increase the number of tasks completed by 12-18\%. 
However, our study also shows that proactive assistants need to be designed carefully as small changes can lead to differing user experiences---for example, increasing the frequency of suggestions can negatively impact user experience, reducing participant preference for the proactive assistant over baseline by half, despite productivity gains.
Additionally, small changes in proactive assistant capabilities, including the ability to integrate suggestions, can significantly change user interaction and coding patterns. 

In summary, this paper contributes:
\begin{itemize}[topsep=0pt]
    \item A proactive chat assistant system, comprising an interactive interface, suggestion content based on the user's code, and logic to appropriately time suggestions (Section~\ref{sec:system}). We also outline a set of design considerations when building proactive assistants (Section~\ref{sec:design_consider} and Section~\ref{subsec:discussion_design}).
    \item A controlled user study characterizing the benefits of our proactive chat assistant on downstream user metrics (e.g., productivity and user experience), highlighting the potential for proactivity to be included in future coding assistants, as well as demonstrating the potential negative effects of changing certain proactive features (Section~\ref{subsec:RQ1} and Section~\ref{subsec:RQ2}). We also analyze fine-grain user behavior when interacting with different proactive assistants (Section~\ref{subsec:RQ3}).
\end{itemize}

\section{Related Work}

\subsection{The Ecosystem of Programming Assistants}\label{subsec:programming_asst}

A growing number of tools powered by large language models (LLMs)---i.e., programming assistants---are available to developers to generate or edit code and answer queries. Programmers are increasingly writing code with AI assistants like Github Copilot~\cite{copilot} and Cursor~\cite{cursor} and are using chat assistants like ChatGPT~\cite{chatgpt} or Claude~\cite{claude} in place of online Q\&A communities like Stack Overflow~\cite{xiao2023devgpt}. 
Programming assistants typically involve one of two types of support: autocomplete suggestions are used to quickly write more code based on the programmer's current code context, while chat dialogue can help answer questions that range from debugging errors to explaining documentation.
Programming assistants surrounding code completion have been the focus of prior work, providing insight into how people use LLM-generated code completions~\citep{vaithilingam2022expectation,mozannar2022reading,barke2022grounded,prather2023its,peng2023impact,cui2024productivity} and improving completions by providing explanations~\cite{yan2024ivie} and determining when to best show~\cite{vasconcelos2023generation,mozannar2024show}.
While there has been less focus on investigating the use of chat assistants for programming, it is important to note that they complement assistance provided by code completions~\cite{liang2023large} and are the focus of this work.

Prior studies investigating chat dialogue for programming assistance have always looked at the setting where people initiate questions to the assistant~\citep{ross2023programmer, chopra2023conversational, kazemitabaar2023studying,xiao2023devgpt,nam2024using,kazemitabaar2024codeaid}.
In particular, \citet{chopra2023conversational} found that 
people spent a significant amount of time constructing prompts and gathering and expressing their context to ChatGPT and did not enjoy being ``slowed-down'' through this process. 
Additionally,~\citet{nam2024using} noted that people often had a hard time finding a good prompt for the chat assistant that could give them the desired response.
Further,~\citet{mozannar2024realhumaneval} highlighted how the burden remains for programmers to appropriately incorporate an assistant response into their code.
These observations highlight the hurdles to a streamlined integration of chat assistants like ChatGPT in a developer's day-to-day workflow: the need to provide sufficient context about the developer's problem and the need for developers to manually input and decide what they want to ask and then act upon the assistant response. 
Even while some programming assistants (e.g., Cursor~\cite{cursor}, Github Copilot~\cite{copilot}) have begun building chat assistants into the integrated development environment (IDE), they still require human input.
In this work, we evaluate whether incorporating an aspect of proactivity into chat assistants can lead to more benefits in terms of productivity and user experience.
Our work aligns with existing tools for programming that provide chat support in that we still allow developers the option to ask questions when they so choose. 
Different from prior work, our proactive chat assistant also periodically recommends suggestions and even allows developers to integrate those suggestions into their code.

While the focus of this work is on designing proactive \emph{chat-based} assistants, we overview other forms of proactive assistance in the existing programming literature. 
As mentioned in Section~\ref{sec:intro} and discussed extensively in Section~\ref{subsec:programming_asst}, the most well-known and widely used form of proactive assistance is in-line code completions (e.g., GitHub Copilot~\cite{copilot})~\citep{liang2023large}.
Aside from proactive code completions, multiple related works have studied proactive assistance to address various issues including fixing error messages and understanding LLM generations.
For example, a set of more traditional approaches explored the use of adaptive feedback---drawn from existing examples or code repair tools---to help students resolve error messages in submitted programs~\cite{errormessage, feedback, ahmed2020characterizing}.
Recent approaches have explored live programming techniques to ``preview''  the outcomes of AI-generated code using projection boxes~\citep{liveprogramming1,liveprogramming}.
Our work investigates whether LLM-powered proactive assistants can address multiple issues at once. 
We return to these studies and other aforementioned studies on chat-based assistance for programming to discuss how our findings inform the design of proactive programming assistants in Section~\ref{subsec:comparison}.

\subsection{Background on Proactive Assistants}

Research on proactive assistants has appeared in many forms, ranging from physical robots~\cite{zhang2015human, baraglia2016initiative,peng2019design} to virtual chat assistants on phones or computer applications~\cite{liao2016can,fitzpatrick2017delivering,gu2023analysts,liu2024compeer}.
Given our target application of programming, we focus our discussion on chat assistants.
Even within proactive chat assistants, the goals of prior proactive assistants vary depending on the context they are used.
For example, support-based assistants use proactivity to provide motivation and continued dialogue~\cite{fitzpatrick2017delivering,liu2024compeer}, while education-related assistants can proactively offer relevant support or explanations for students~\cite{winkler2019bringing}.
Most relevant to our domain are the set of information-finding assistants that leverage proactive messaging to provide additional and useful information~\cite{avula2018searchbots,hu2024designing} and planning assistants can proactively help people reason about their decisions and identify overlooked alternative decisions or rationales~\cite{gu2023analysts}.
The goal of our proactive assistant is to improve a programmer's productivity while maintaining a good user experience; it requires blending multiple goals of existing proactive assistants.
In Section~\ref{sec:design_consider}, we discuss further how these downstream metrics inform the design considerations for our proactive assistant.

Many prior deployed proactive assistants have failed or received significant negative reaction because the actual capabilities of these systems do not meet user expectations~\cite{luger2016like,meurisch2020exploring}, as many of the systems relied on pre-set messaging or simple models~\cite{liao2016can,fitzpatrick2017delivering}.
With the advent of modern LLMs, and their growing usage as the backbone of agents~\cite{zhou2023webarena,park2023generative,yang2024swe}, there is significant potential to revisit how we design more capable proactive assistants. 
In recent work, ComPeer~\cite{liu2024compeer}  leveraged GPT-4 to build a proactive assistant for mental health support that handles and acts on a longer term memory of user dialogue.
Aside from mental health applications, we believe LLMs can be particularly well suited for proactive assistants for productivity goals, particularly when they are already being used as regular chat assistants (e.g., ChatGPT).
The increased model capability and ability to handle long contexts would be suitable for capturing complex environments in which they are completing tasks, which has been relatively unexplored in prior chat assistant literature relating to productivity~\cite{chaves2021should}.
In Section~\ref{sec:system}, we discuss the design of our proactive system which leverages off-the-shelf state-of-the-art LLMs to propose suggestions based on the user environment and even incorporate these suggestions into user code.

\section{Design Considerations for Proactive Coding Assistants}\label{sec:design_consider}


To formulate design considerations for the core functionality and behavior of the proactive assistant, we revisit existing guidelines on designing mixed-initiative user interfaces and human-AI interactions~\cite{horvitz1999principles,amershi2019guidelines} as well as a survey of known challenges of designing proactive chat assistants~\cite{chaves2021should}.
We highlight five design considerations that merit further attention to capture the benefits and challenges of proactivity.

\textbf{Harnessing the benefits of proactivity.} 
 \citet{horvitz1999principles} emphasizes the importance of ``developing significant value-added automation'' when developing effective mixed-initiative systems.
Given prior work highlighting the limitations of existing chat assistants for coding contexts~\cite{chopra2023conversational, nam2024using, mozannar2024realhumaneval} (as discussed in Section~\ref{subsec:programming_asst}), the value-add of introducing proactivity would be an improvement in programmer productivity (e.g., in terms of time and effort) and in the user experience (e.g., reducing perceived ``slow-down''), as compared to a non-proactive chat assistant.
Concretely, we propose the following two design considerations:

\textit{(1) Support efficient evaluation.} 
Proactive suggestions should be provided in a way such that programmers can efficiently interact with them. This involves providing a manageable amount of information to the user about the suggestions so that they can quickly decide if they are worthwhile.
This might involve hierarchically scaffolding the suggestions so that the programmer can obtain more information about the suggestion if needed.

\textit{(2) Support efficient utilization.} If the user decides the suggestion is useful, the proactive assistant should make it easy for the programmer to utilize it.
For example, the proactive assistant should make it as easy as possible for them to take action corresponding to the suggestion and incorporate it into the programmer's code.
If the programmer decides the suggestion is not useful, they should be able to easily dismiss the assistant.

\textbf{Handling challenges of proactivity.} Prior works on proactive chat assistants have demonstrated that there is considerable nuance in developing a proactive chat assistant~\cite{chaves2021should}, as untimely and irrelevant proactive messages may compromise the success of the interaction timing~\cite{silvervarg2013iterative, liao2016can,toxtli2018understanding,tallyn2018ethnobot}. 
We incorporate these challenges in our design considerations:

\textit{(3) Show contextually relevant suggestions.} Proactive chat assistants are increasingly built into complex environments where users often perform a number of different tasks. 
For example, programming tasks can range from implementation to debugging and testing, just to name a few, which all would require different interventions and information from the assistant. 
Proactive assistants should make sure to surface information that is relevant to the current context and recent user interactions and messages.

\textit{(4) Incorporating user feedback.} The proactive assistant should adapt its suggestions and timing as it interacts with the user. The user should be able to accept and reject proactive suggestions, and the decision to accept or reject suggestions should influence future decisions of the proactive assistant. This is especially important as different programmers may want different levels of proactive help from the assistant based on their experience for instance.

\textit{(5) Time suggestions based on context.} 
Programmers perform various tasks during their work and are often in a state of flow \cite{csikszentmihalyi2014flow}. The proactive assistant should consider what the programmer is currently doing and their workflow before it provides a suggestion \cite{barke2022grounded}. For example, showing suggestions while the user is actively coding can disrupt the programmer's flow and lead to disruptions.
\citet{horvitz1999principles} discusses the importance of fallback behavior which can allow for a reduced frequency of suggestions.


In the following section, we introduce an implementation of a proactive assistant for programming by building on the design considerations above.


\section{Proactive Assistant Implementation} \label{sec:system}


In this paper, we build a proactive assistant for programming tasks based on the design considerations in Section \ref{sec:design_consider}. 
Our proactive chat assistant distinguishes itself from online AI assistants such as ChatGPT by first having access to the user's work context and second by proactively proposing suggestions to the user versus only passively waiting for user requests.
When embedding the proactive assistant into the user's work context, which is the programmer's IDE in this setting, the proactive assistant will have access to the current code in any of the user's files, the terminal outputs, and prior user queries to the AI assistant.
To turn the context into suggestions, we discuss different components that comprise our proactive assistant and how we instantiate each design consideration, as shown in Figure~\ref{fig:system_overview}: 
\begin{itemize}
    \item \textit{Interactive interface}: In Section~\ref{subsec:interface}, we overview the interface through which the assistant presents the generated suggestions in a way that makes it easy for users to interact with, understand, and invoke when needed (Design Considerations 1-2).
    \item \textit{Suggestion generation}: In Section~\ref{subsec:suggest_gen}, we discuss how the assistant takes the current context and generates a set of proactive suggestions that is timely and relevant (Design Consideration 3) which take into account user feedback (Design Consideration 4). In Section~\ref{subsec:timing}, we discuss how to appropriately determine when suggestions are shown to users (Design Consideration 5). 
\end{itemize}

\subsection{Interactive Interface} \label{subsec:interface}

\begin{figure*}[t]
\centering
\includegraphics[width=\textwidth]{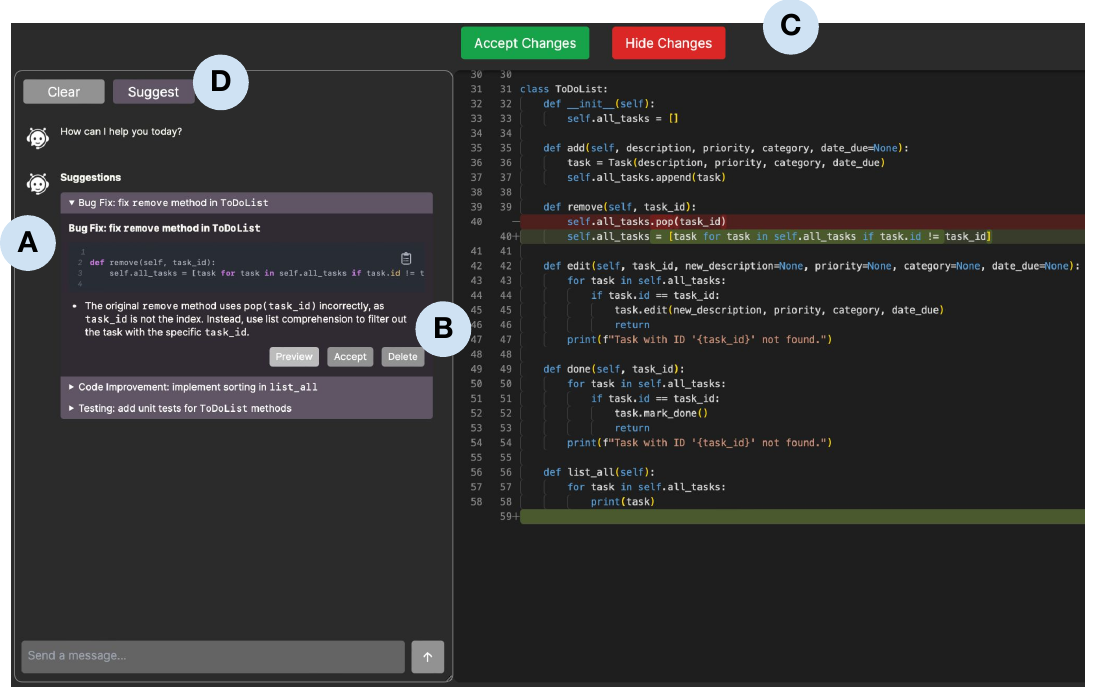}
\caption{\textbf{A walkthrough of the proactive assistant interface for coding.} (A) Overview of suggestions, where the assistant provides a short description, and users can choose to expand a suggestion for more details (this can involve implementation and a brief explanation); (B) Buttons that allow participants to preview the implementation in their code, accept a suggestion to ask follow-up questions on, or delete the suggestion; (C) Integrating a suggestion into the editor via a diff format, where users can decide if they want to accept or hide the changes; (D) Invoking suggestions, where users can also request suggestions from the assistant (for example, if they did not like the original suggestion or wanted a suggestion before the assistant provided one).}
\Description{\textbf{A walkthrough of the proactive assistant interface for coding.} (A) Overview of suggestions, where the assistant provides a short description, and users can choose to expand a suggestion for more details (this can involve implementation and a brief explanation); (B) Buttons that allow participants to preview the implementation in their code, accept a suggestion to ask follow-up questions on, or delete the suggestion; (C) Integrating a suggestion into the editor via a diff format, where users can decide if they want to accept or hide the changes; (D) Invoking suggestions, where users can also request suggestions from the assistant (for example, if they did not like the original suggestion or wanted a suggestion before the assistant provided one).}
\label{fig:interface_overview}
\end{figure*}

The interface of the proactive assistant, shown in Figure~\ref{fig:interface_overview}, is built into the standard chat interface, allowing programmers to use all normal chat functionality, while the assistant periodically provides suggestions to programmers.
Furthermore, this means the proactive assistant does not take up additional real estate on the screen, particularly when programmers may already have multiple files open side-by-side.
We describe the different features of the proactive suggestion and their respective design considerations:

\subsubsection{Suggestion summary.} Towards fulfilling design consideration 1, the proactive assistant facilitates efficient evaluation of the suggestion by providing a summary to allow users to quickly get an idea of whether or not the suggestion is relevant. The summary consists of a single sentence that starts with the type of the suggestion e.g., a bug fix, a new feature, and then a description of the suggestion, as shown in the headings of  Figure~\ref{fig:interface_overview} (A). 
The drop-down style of the interface allows users to easily expand and condense a suggestion and quickly browse the various suggestions.

\subsubsection{Details of suggested implementation.} 
If the summary of the suggestion seems relevant to the user, they can expand the suggestion for more details as shown in Figure~\ref{fig:interface_overview} (A).
Depending on the type of suggestion, they may receive a code snippet and/or a text description (e.g., a suggestion that explains a code snippet may not include code).
The interface also allows users to copy the suggestion into the editor (design consideration 2).
We limit the text description of the implementation to a few bullet points to facilitate efficient interaction (design consideration 1).

\subsubsection{Preview suggested implementation in code.} While users have the freedom to copy the suggestion code into the editor and decide how they want to use the suggestion, the proactive assistant also has a ``preview'' functionality, which will allow users to see how the assistant would incorporate it into their code (design consideration 2). 
The suggested implementation is computed via another call to the LLM, where it is given the proactive suggestion and the user's code and produces a new code that incorporates the suggestion. 
Each time a suggestion is previewed requires another LLM inference call. As such, we only preview a suggestion when invoked by the user, rather than automatically previewing \emph{every} suggestion, which helps to reduce the latency that users experience.
This new code file generated is then shown to the user via a diff editor based on the Monaco Editor \footnote{\url{https://microsoft.github.io/monaco-editor/}} which tells the user what lines were added or removed once the suggestion was incorporated, as shown in Figure~\ref{fig:interface_overview} (C).
The user has the option to accept or revert all changes and can additionally only a subset of the changes and edit freely before clicking on the ``accept changes'' button.
In the Supplementary Material, we include the prompt used to incorporate edits into the current code snippet.

\subsubsection{Asking follow-up questions based on suggestion.} Since the proactive assistant is a part of the chat interface, we create the option for users to add a proactive suggestion to the chat history (via the ``accept'' button) or remove it (via the ``delete'' button). Clicking on the accept button shown in Figure~\ref{fig:interface_overview} (B) would add the expanded suggestion to the chat message in a way that looks visually similar to the rest of the user's messages. This allows future proactive suggestions to condition on the user's accept or delete actions addressing design consideration 4 (incorporating user feedback). This also allows the user to conveniently ask follow-up questions on the proactive suggestion. This may be helpful as it is common for users to follow up on LLM responses to clarify or correct the model.

\subsubsection{Manually requesting suggestions.} While the nature of a proactive assistant suggests that the assistant will be the one who determines when to provide suggestions to the user, it is unrealistic to expect an assistant always to anticipate when users might want suggestions (design consideration 5). 
As such, we include a button, as shown in Figure~\ref{fig:interface_overview} (D), allowing users to request suggestions as they wish. This can also allow us in future work to better learn when the proactive assistant should intervene.

\subsection{Suggestion Content}\label{subsec:suggest_gen}

\begin{table*}[t]

\setlength{\tabcolsep}{3pt}
\centering
\label{tab:example_questions}
\begin{tabular}{@{}p{0.25\textwidth}  p{0.75\textwidth} @{}}
\toprule
\textbf{Suggestion Type} & \textbf{Example of developer question} \\
\midrule \midrule
Explaining existing code
& 
What does this do? \newline
\texttt{model = GPTLanguageModel()}\newline
\texttt{m = model.to(device)}\\
\midrule
Brainstorming new\newline
functionality &
Based on the following OCaml code for an s-expression evaluator, write a parser for the tokens defined. Note that all operations must be binary and bracketed, with no concept of operator precedence.\newline
\texttt{[code context]}\\
\midrule
Completing unfinished code &
The following is some C code for binding a "hello" C function to an Io method. Can you complete the code by the comment:\newline
\texttt{[code context]}\\
\midrule
Pointers to documentation & 
I've saved this in `scripts/align-import.py'.\newline 
Remind me how to use `find' to run it on every `*.lean` file in a subdirectory?\\
\midrule 
Debugging \newline
(Latent errors) &
Why is my redirect not working? Here is my client side code\newline
\texttt{[code context]}\newline
Everything else works as intended, except that it will not redirect. What is the issue here?\\
\midrule
Debugging \newline
(Runtime errors) &
\texttt{[code context]}\newline
Test suite failed to run\newline
Cannot find module `libp2p' from
`src/shared/libp2p\_node/index.ts'\newline
\texttt{[remainder of error trace]}\newline
Do you have any idea why I am getting this error?\\
\midrule
Adding unit tests &
Could you create Jest unit tests for this function?\newline
\texttt{[code context]}\\
\midrule
Improving efficiency and \newline
modularity & 
The following is a kernel of an algorithm. It uses Apple’s metal api for matrix operation. I think it can be improved to make it run faster.\newline
\texttt{[code context]}\\
\bottomrule
\end{tabular}
\caption{\textbf{Suggestion types informed by dataset analysis.} We identify categories of suggestions that a proactive assistant can surface by coding developer questions to ChatGPT in the DevGPT~\cite{xiao2023devgpt} dataset. We incorporate these suggestion types into the design of our proactive assistant.} 
\Description{\textbf{Suggestion types informed by dataset analysis.} We identify categories of suggestions that a proactive assistant can surface by coding developer questions to ChatGPT in the DevGPT~\cite{xiao2023devgpt} dataset. We incorporate these suggestion types into the design of our proactive assistant.}
\label{tab:examples}
\end{table*}

For the proactive assistant to be useful, it must generate appropriate suggestion content (design consideration 3).
This means that the assistant should have knowledge of the types of suggestions that users benefit from and that the assistant should surface the right type of suggestion at the right time, depending on the user's code context.

To identify the different types of suggestions that users benefit from for coding contexts, we look to DevGPT~\cite{xiao2023devgpt}, a dataset of conversations between programmers and ChatGPT, as a source of real-world questions.
While other chat datasets also contain in-the-wild programming-related questions (e.g., WildChat~\cite{zhao2024wildchat}, LMSYS-Chat-1M~\cite{zheng2023lmsyschat1m}), we selected DevGPT in particular because the majority or all questions contain user code context in addition to the user question and LLM response, which allows us to manually inspect whether we think it would be reasonable for a proactive assistant to anticipate this question from the current code context.
We sample 100 questions and their code contexts from the DevGPT data and manually create a label for the ``category'' of the question.
We then cluster labels to identify common themes across questions.
The goal of surveying existing questions is to ensure we obtain a reasonable set of potential suggestion types to prompt the proactive assistant, however, this set may not be exhaustive.

From this annotation process, we identify eight categories of questions that proactive assistants can help with: \emph{explaining existing code}, \emph{brainstorming new ideas or functionality}, \emph{completing unfinished code}, \emph{providing pointers to syntax hints or external documentation}, \emph{identifying and fixing bugs} (which include both latent and runtime errors), \emph{adding unit tests}, and \emph{improving code efficiency and modularity}.
Annotations were roughly distributed across all categories where each appeared 15\%, 19\%, 18\%, 13\%, 10\%, 9\%, 6\%, 10\% of the time respectively---note that since this dataset consists of only ChatGPT conversations and may not represent the actual distribution of user questions in practice.
In Table~\ref{tab:examples}, we include example user questions that fall under each category.

\subsection{Timing of Suggestions}\label{subsec:timing}

The proactive assistant must display suggestions at the appropriate time so they are not obstructive and distracting to users (design consideration 5).
A study by~\citet{barke2022grounded} characterized that interaction with programming assistants consists of two modes: acceleration,
where the assistant is primarily used to help the programmer implement what they already intend to write; and exploration, where the programming assistant assists a programmer in identifying and planning out goals.
We propose the following conditions for when proactive suggestions should be shown to users based on an estimation of when the user is in acceleration or exploration mode: 
\begin{itemize}
    \item \textit{During acceleration:} Suggestions are not requested when the user is interacting with a previous suggestion, typing in the chat, sending a message, or waiting for a response. The suggestion timer resumes 5 seconds after
    the user stops typing to account for a typing break. If the user would like a suggestion, they can still request it.
    \item \textit{During exploration:} When the user is idle—likely due to planning and brainstorming, the assistant provides
    suggestions after 5 seconds, limited to every 20 seconds since the last suggestion, interaction, or chat. If the user starts coding before suggestions are shown, due to potential latency in the model query, we do not display the suggestions. 
\end{itemize}
\noindent To detect mode based on user interactions, we implement listeners and corresponding timers in the editor to track whether the user is actively taking actions (e.g., writing code or chat messages). 
Based on pilot studies with 4 participants, we determined that 5 seconds was a reasonable estimates of how long people tend to pause when coding and 20 seconds was a reasonable guess at how much time was needed to check out the proactive suggestions.

Additionally, we allow the proactive suggestion to be shown \textit{during debugging} when users are trying to edit the current code for any bugs or improve the performance.
Debugging is distinct from the previous two states because the user is neither trying to write more code nor planning the next steps. 
The proactive assistant immediately provides a suggestion when the user runs or submits code that leads to an error because the user is not actively coding.

\begin{figure*}[t]
\centering
\includegraphics[width=0.8\textwidth]{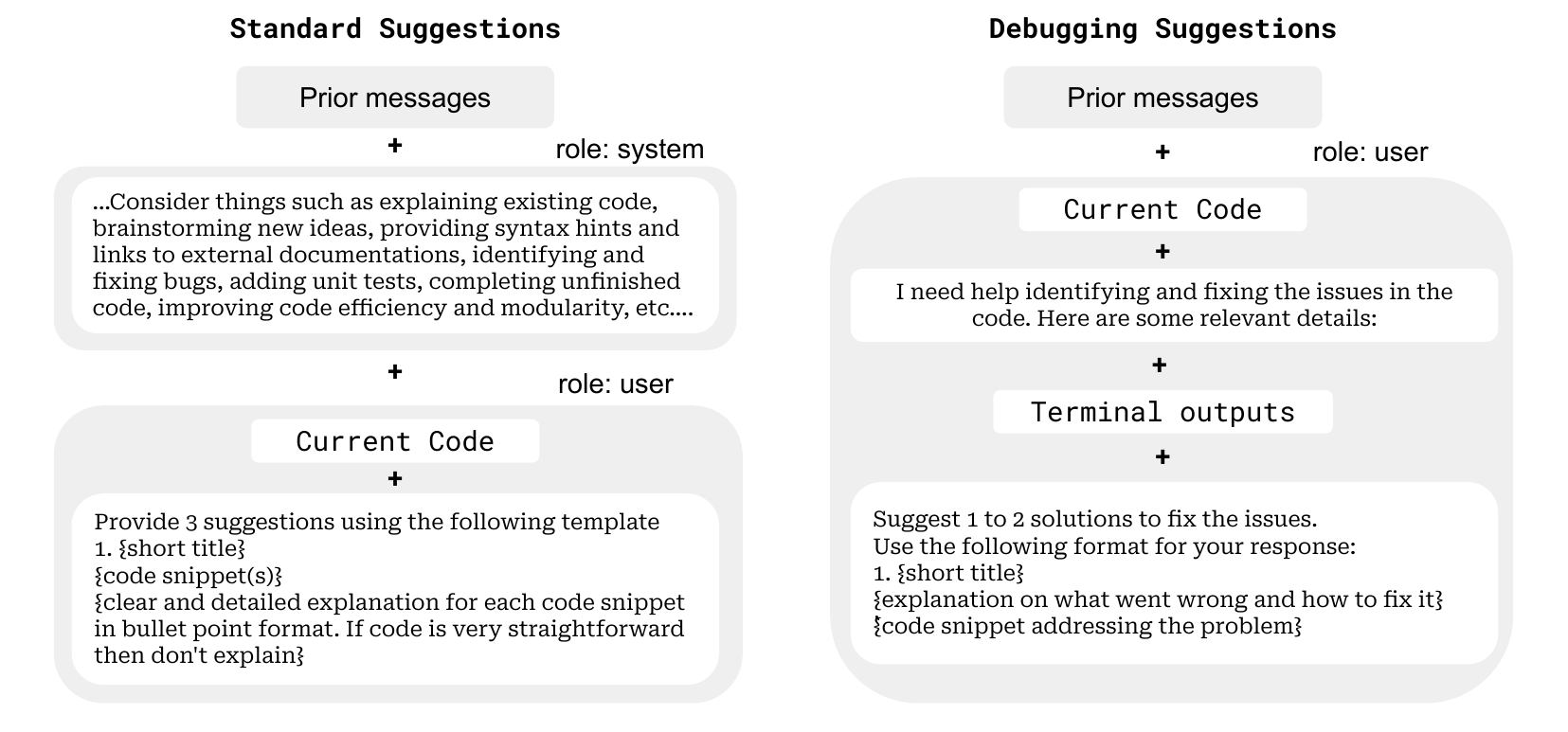}
\caption{\textbf{Prompting the assistant for suggestions.} An overview of how different inputs (e.g., prior messages, current code, and terminal outputs) are incorporated in the generation of standard and debugging suggestions, the former which is shown generally and the latter which is only triggered when code is run. Full prompts for both types of suggestions are provided in the Supplementary Material.}
\Description{\textbf{Prompting the assistant for suggestions.} An overview of how different inputs (e.g., prior messages, current code, and terminal outputs) are incorporated in the generation of standard and debugging suggestions, the former which is shown generally and the latter which is only triggered when code is run. Full prompts for both types of suggestions are provided in the Supplementary Material.}
\label{fig:prompts}
\end{figure*}

\subsection{Generating Suggestions} \label{subsec:llm_call}

To generate a set of suggestions, we discuss how the assistant makes a call to the LLM with the current user context to generate a set of suggestions.
In this work, all LLM calls are made to GPT-4o \cite{chatgpt}, a state-of-the-art LLM. 
In pilot studies, we experimented with smaller, faster models (e.g., GPT-4o-mini) but observed that the quality (e.g., correctness of suggested code, relevance of suggestions) of suggestions to be lower.
We overview what inputs from the user context are used to generate each type of suggestion and include the full prompts in the Supplementary Material.

\paragraph{Standard suggestions.} 
Standard suggestions are the ones that periodically show up while the user is writing code.
To generate these suggestions, the LLM call includes a list of items: any prior chat messages, a system prompt, and a general prompt. 
We include the prior messages to avoid generating suggestions that may repeat previously asked questions and to make the model aware of prior user-assistant interactions. 
Then, we scaffold the suggestion types identified in Section~\ref{subsec:suggest_gen} into the system prompt to encourage the model to consider a more diverse set of suggestions. 
Finally, we include the current code and instructions for how to return the set of generated suggestions in the main prompt. Figure~\ref{fig:prompts} (left) shows how all inputs are combined to generate standard suggestions.

\paragraph{Debugging suggestions.} Debugging suggestions are the suggestions that are only triggered when the user runs their code. 
The LLM call again includes prior messages, but instead of the system prompt, we simply include all other information in one main prompt.
This main prompt includes the current code, a short instruction to ask the model to focus on any error outputs, error statements from the terminal, and instructions for how to return the set of generated suggestions.
Note that debugging suggestions have a more restricted set of suggestions that the assistant can generate. Figure~\ref{fig:prompts} (right) shows how all inputs are combined to generate debugging suggestions.

The response from the LLM will then be parsed into individual suggestions and displayed in the exact order in which they are returned in the interactive interface as described in Section~\ref{subsec:interface}.
While we do not explicitly incorporate an additional mechanism to rank suggestions, we verify based through pilot tests that the order of suggestion types is not always the same.

\section{Study Design} \label{sec:study_design}

We conducted a controlled user study to understand the effectiveness of a proactive assistant on productivity and investigate how people interact with the assistant.
Our research questions are as follows:
\begin{itemize}
    \item[\textbf{RQ1:}] \textit{What is the effect of the proactive assistant on user productivity? }The goal of a proactive chat assistant is to anticipate potential user needs or even surface questions the user has not considered. As such, we expect the assistant to have a measurable downstream effect on productivity measures.
    \item[\textbf{RQ2:}] \textit{What is the effect of the proactive assistant on user experience?} In addition to productivity measures, we hope the user has a favorable experience when interacting with proactive assistants.
    \item [\textbf{RQ3:}] \textit{How do participants interact with different proactive assistants?} A proactive assistant comprises several design considerations, we analyze how users interact with different variants of our proactive assistant to understand the effect of these design decisions on user behavior. 

\end{itemize}

\subsection{Participants}

We recruited a total of 65 students via university mailing lists. The inclusion criteria for the study are that they must be based in the U.S., be older than 18 years old, and have experience programming. 
Since we design tasks that are written in Python, we require participants to have a baseline level of Python knowledge. 
Among our participants, 60\% reported their Python knowledge as intermediate, 31\% had advanced Python experience, and 9\% were beginners. Moreover,
19\% self-report that they are daily users of AI tools for programming (including GitHub Copilot and ChatGPT), 46\% report that they use these tools at least once if not multiple times a week, and 35\% rarely use AI tools for programming.  
The gender breakdown of our participants was 34\% female and 66\% male. The student population was distributed between 38\% undergraduate and 62\% graduate students, skewing towards more experienced programmers.

\subsection{Baseline}

Our study compares proactive assistants to a non-proactive chat-based assistant baseline. 
Every participant in our study interacts with a standard chat assistant that is incorporated into the coding environment but does not have access to the user's code. This is similar to the set-up that has been studied in prior papers~\cite{kazemitabaar2024codeaid,xiao2023devgpt, mozannar2024realhumaneval} and mimics how people would use ChatGPT. 
The interaction paradigm for the baseline is as follows: the user needs to type in their query to obtain a response from the chat assistant. The chat assistant can respond in natural language and provide code snippets. 

In the baseline (as well as our experimental conditions), we do not include any code autocomplete tools to not introduce a potential confounding factor in our analysis as not all participants are experienced in using autocomplete tools.
Further, we believe the proactive assistant can complement code autocompletion tools, which consist of local proactive changes, while the proactive assistant provides suggestions at a more global level.


\subsection{Experimental Conditions}\label{subsec:exp_conditions}

We consider three experimental conditions to explore the effect of different proactive chat assistant designs:

\paragraph{Condition 1: Our proactive chat assistant (\texttt{Suggest and Preview}):} In this condition, the participant interacts with the proactive assistant described in Section~\ref{sec:system}. This condition encompasses all features of the baseline condition, i.e., the participant still has the option to send chat messages to the assistant at any point while also receiving proactive suggestions.

\paragraph{Condition 2: Our proactive chat assistant without preview feature  (\texttt{Suggest}):} In this condition, the participant interacts with the proactive assistant described in Section~\ref{sec:system}, without the preview feature, which allows the assistant to suggest direct edits to the user's code.
Similar to Condition 1, the participant can still send chat messages to the assistant at any point while also receiving proactive suggestions.

\paragraph{Condition 3: Persistent proactive chat assistant  (\texttt{Persistent Suggest}):} 
In this condition, the participant interacts with a variant of our proactive chat assistant. 
While the proactive assistant in this condition looks at face value to be the same as the proactive assistant in Condition 2 (\texttt{Suggest}), we modified a few different parameters discussed in Section~\ref{sec:system}: reducing the amount of time that an assistant waits to provide a suggestion (from 20 seconds to 5 seconds), increasing the number of suggestions shown (from 3 to 5), and removing guiding prompts. 

The three conditions allow us to investigate whether (1) the level of proactivity affects user productivity and user experience (comparing \texttt{Suggest and Preview} and \texttt{Suggest}) and whether (2) small changes in the design of a proactive assistant can make a sizeable difference in user productivity and experience (comparing \texttt{Suggest} and \texttt{Persistent Suggest}). 
Recall we fix the LLM backbone across all conditions to be GPT-4o \cite{chatgpt}, a state-of-the-art LLM.

\subsection{Experimental Platform}

The study was conducted online and asynchronously at the participant's own time. 
The interface we used to deploy the study is an extension of the open-source platform  RealHumanEval introduced in~\cite{mozannar2024realhumaneval}.
Conducting the study on the web allows for easy assignment of participants to experimental conditions, the duration of the study, and the tasks participants solve.
Finally, the web interface reduces installation issues or any potential incompatibility. 
While the web interface is not the traditional IDE that a user may typically use for their day-to-day development, it is reminiscent of online coding platforms (e.g., Leetcode) that participants are generally familiar with.
The editor in our web interface is also the same Monaco editor as in Visual Studio Code, a popular IDE.
We also added a check in the post-study form to make sure that participants did not have any issues using the interface.

\subsection{Procedure}

Before participating in the study, each participant filled out a consent form.
The study has been approved by our institution's review board (IRB).
We adopt a between-subject setup to compare the three proactive chat variants and a within-subjects setup to compare each proactive variant to the non-proactive baseline.
This means each participant will interact with both a proactive assistant and the baseline chat condition.
We randomize the order in which participants interact with either a proactive assistant or the baseline chat assistant. 
The total amount of time the participant spent coding for the study was 40 minutes, with participants spending 20 minutes in each condition. 

\subsubsection{Onboarding}
Before participants can access the proactive assistant, they are provided with a short tutorial that describes how to interact with the proactive suggestions.
The tutorial highlights how the proactive assistant will provide a high-level summary of suggestions that can be expanded upon and how to ask follow-up questions.
The tutorial also warns participants that the suggested implementations may not always be correct and to carefully verify the suggestions.
In the Appendix, we provide a copy of the onboarding instructions that participants had access to.

\subsubsection{Coding Tasks} 
In our study, participants engage in 20-minute sessions focusing on two types of programming problems designed to reflect real-world engineering scenarios. The first type of problem aims to test system-building skills. For instance, participants are tasked with enhancing an ``online store" class object implementation. Given an initial starter code, they must implement additional features specified through natural language instructions, brainstorm a new feature, debug potential issues, and write tests for these new features. The second type of problem challenges participants to work with unfamiliar packages, simulating situations where engineers must quickly adapt to new tools. An example of this is manipulating a dataframe using various NumPy functions. Unlike previous studies that primarily assessed puzzle-solving abilities through Leetcode-style problems \cite{vaithilingam2022expectation,mozannar2024realhumaneval}, our approach aims to evaluate skills more directly relevant to practical software engineering tasks.
In the Appendix, we provide task descriptions for each of the tasks used in the study.

The problem types are randomized across conditions and in both proactive/baseline conditions to avoid confounding task difficulty or type with the helpfulness of proactive assistants.
Since each participant experienced both the baseline and proactive conditions, we distributed the 4 tasks in a symmetrical manner.
This means that if the participant started with a system-building task, they would see the other system-building task when the proactive assistant was switched on or off. 
This scheme also ensures a relatively similar number of task types in both proactive and baseline conditions.
Participants decide when they are satisfied with their current attempt on the given task and would like to move on to another question.
As such, we had a second question of the other task type queued up for the participant to work on for the remainder of the time within that condition.
In sum, each participant attempted two tasks of the same type across the baseline and proactive conditions, and \emph{additionally} one or two of the remaining tasks of the different task type when time permitted.

\subsubsection{Post-task survey} 
The study concluded with a post-task survey that asked participants about their experience with both the proactive and baseline conditions. While we log telemetry data in both conditions to measure user productivity, we use the post-task survey to measure user perceptions.
In the Appendix, we provide a list of the post-task survey questions.

\subsection{Measurements}

To answer our research questions, we measured the following: 
\begin{enumerate}
    \item \textit{Telemetry.} Our interface logs all user behavior from coding to using the chat assistant: anytime the code is updated by the user, the interface saves the updated code. Further, any chat messages and associated responses are logged. Finally, any interactions with the proactive assistant, including expanding suggestions, accepting a suggestion, and requesting suggestions are all recorded.
    Using the collected telemetry data, we compute the the number of sub-tasks completed.
    We can also obtain fine-grained interaction behavior to identify common interaction patterns.
    \item \textit{Perceived usefulness.} In the post-task survey, we ask participants to rate their interactions with both the proactive and baseline chat assistants. Participants were then asked to explain their ratings.
    \item \textit{Open-ended feedback.} In the post-task survey, we also asked additional questions that further dive into what aspects of the proactive assistant participants preferred and what could be improved.
\end{enumerate}

\subsection{Analysis Approach}

To measure the effect of proactive assistants on user productivity (\textbf{RQ1}), we computed the number of sub-tasks completed.
We build a linear model which incorporates the experimental condition, which is either baseline or one of the proactive models, and the coding problem as fixed effects.
We also qualitatively inspect whether the code written contains automated test cases.
To understand the effect of proactive assistants on user experience (\textbf{RQ2}), we measure participant ratings of the baseline and proactive assistant.
We then computed a binary measure of whether participants preferred the proactive assistant to the baseline and ran an ANOVA with Tukey HSD.
We also analyze participant justifications to interpret the quantitative results.
For all tests, the threshold for statistical significance was $\alpha=0.05$.
To begin to understand how participants use the different proactive assistants (\textbf{RQ3}), we measure the frequency of interactions with the various components of the proactive assistants as logged in the telemetry.
We consider this portion of the results an exploratory analysis to identify trends that distinguish proactive conditions.
We present the results of our study in the following section.

\section{Results}\label{sec:results}

\subsection{RQ1: Proactive Assistants Generally Improve User Productivity} \label{subsec:RQ1}

\begin{figure*}[t]
\centering
\includegraphics[width=0.8\textwidth]{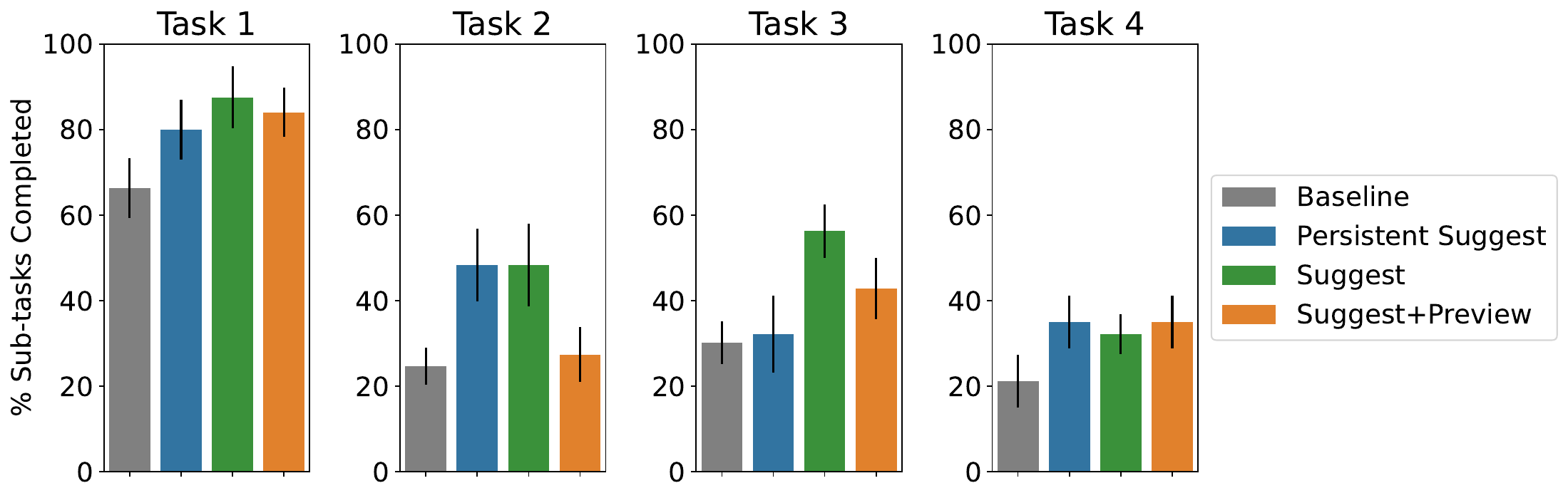}
\caption{\textbf{Percentage of sub-tasks completed correctly.} Comparing baseline chat to proactive assistants across the four tasks, where Task 1 and Task 2 are system-building questions and Task 3 and 4 are ones where participants work with new packages and functionality. We report average performance and standard error. While performance varied by task, we observed that all variants of proactive assistants tended to increase the number of test cases passed compared to the baseline chat assistant across the board.}
\Description{\textbf{Percentage of test cases passed.} Comparing baseline chat to proactive assistants across the four tasks, where Task 1 and Task 2 are system-building questions and Task 3 and 4 are ones where participants work with new packages and functionality. We report average performance and standard error. While performance varied by task, we observed that all variants of proactive assistants tended to increase the number of test cases passed compared to the baseline chat assistant across the board.}
\label{fig:productivity}
\end{figure*}

\textit{Number of sub-tasks completed.} When comparing the number of sub-tasks completed by participants, we find that on average participants with proactive assistants are more productive with a baseline chat assistant: we observe a $12.1\% \pm 5.1\%$, $18\% \pm 5.8\%$, and $11.6\%\pm 5.0\%$ increase in the percentage of test cases passed for \texttt{Suggest and Preview}, \texttt{Suggest}, and \texttt{Persistent Suggest} respectively compared to baseline.
Figure~\ref{fig:productivity} provides a more granular view by task.
We find that the improvements in the number of test cases are significant across \emph{all} proactive variants, where $p=0.01$ for \texttt{Suggest and Preview}, $p=0.002$ for \texttt{Suggest}: $p=0.002$, and $p=0.02$ for \texttt{Persistent Suggest}.
We explore whether different groups of participants benefit from proactive suggestions differently (Appendix~\ref{sec:additional_results}): we do not observe any significant differences based on Python expertise or AI tool usage frequency but do observe a significant difference based on gender (i.e., women tend to benefit more from proactive assistance).

\textit{Number of test cases written.} In addition to the increase in the number of test cases passed, we also observe an increase in the amount and quality of test cases written. Note that writing tests was part of the instructions provided to users. 
Focusing on Task Type B, where it is important to test how different components of the designed system work together, we annotate user code for whether they incorporated test cases and find that only 13.3\% do in the baseline setting while 33.3\% do in the proactive settings.
We believe that providing proactive suggestions can help create a virtuous cycle where including more test cases helps participants surface issues in their code, thus also increasing the number of test cases passed.

\subsection{RQ2: User Experience with Proactive Assistants Varies by Implementation.} \label{subsec:RQ2}

\begin{figure}[t]
\centering
\includegraphics[width=0.7\columnwidth]{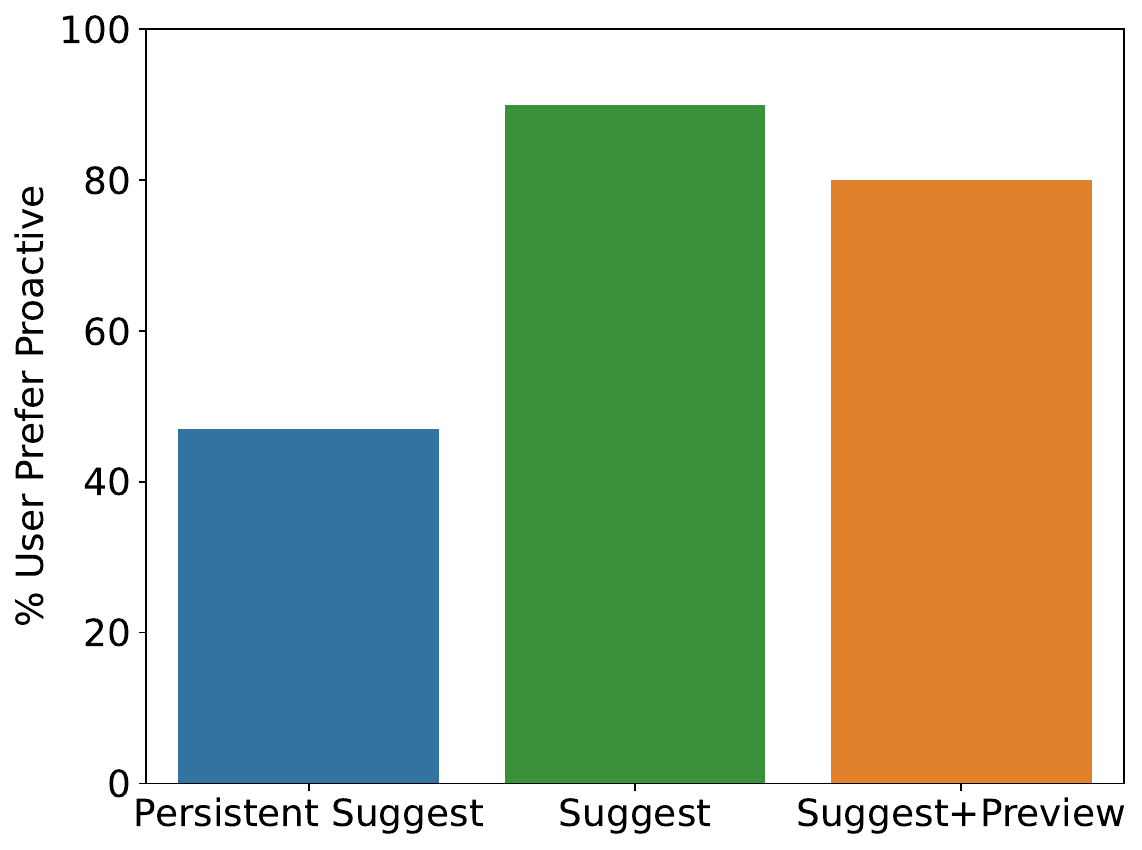}
\caption{\textbf{Comparing participant perception of proactive versus baseline.} User experience is important to the adoption of proactive assistants. We compare how often participants preferred a proactive assistant to the baseline variant in each condition and find that participants generally preferred the proactive variant only in the \texttt{Suggest} and \texttt{Suggest and Preview} conditions.}
\Description{\textbf{Comparing participant perception of proactive versus baseline.} User experience is important to the adoption of proactive assistants. We compare how often participants preferred a proactive assistant to the baseline variant in each condition and find that participants generally preferred the proactive variant only in the \texttt{Suggest} and \texttt{Suggest and Preview} conditions.}
\label{fig:user_experience}
\end{figure}

\textit{User preferences across proactive conditions.} While we observe generally uniform benefits of proactive assistants across the different conditions in terms of user productivity, we see more variation in terms of user experience across the conditions.
As shown in Figure~\ref{fig:user_experience}, we measure whether participants preferred the proactive assistant over the baseline non-proactive assistant. 
We find that the vast majority of participants prefer the proactive variant over the baseline for both the \texttt{Suggest} and \texttt{Suggest and Preview} proactive assistants (90\% and 80\% of participants respectively).
In contrast, less than half of participants (47\%) in the \texttt{Persistent Suggest} proactive condition preferred having the proactive assistant.
We find that both \texttt{Suggest} and \texttt{Suggest and Preview} variants are statistically different than the \texttt{Persistent Suggest} variant ($p=0.005$ and $p=0.03$ respectively).
On average, participants had a neutral view of the baseline assistant in terms of usefulness. 
As such, participants viewed both the \texttt{Suggest} and \texttt{Suggest and Preview} variants on average as net beneficial.

\paragraph{Comparing participant responses across proactive variants.}
Participants in the \texttt{Persistent Suggest} condition tended to not prefer the proactive assistant because they often found the assistant to be \textit{``distracting''} and  \textit{``annoying''}. 
One participant noted that \textit{``the non-proactive chat assistant was best because it didn't interrupt what I was doing.''} and another participant \textit{``found it annoying because it distracted [them] from working with the non-proactive answers.''}
In contrast, participants in the \texttt{Suggest} and \texttt{Suggest and Preview} conditions tended to have a positive view of the suggestions provided by the proactive assistant compared to the baseline.
One participant noted that they preferred the \textit{``non-proactive was pretty useless because it didn't have any of the context regarding what I was trying to do or the task; I could only really use it for pure syntax like what I would usually search StackOverflow for. Proactive was better because it had more context.''} and another stated that \textit{``I really wasn't sure what to ask for with the non-proactive chat''} since the baseline chat \textit{``required manual input to generate advice''}.

Further, we identified two reasons for why participants did not prefer proactive suggestions across proactive conditions. 
The first reason was concerns about irrelevant suggestions, where participants thought \textit{``it was hard to clearly understand what their use cases were''} and \textit{``they did not seem useful to the particular tasks''}.
Interestingly, participants were more concerned about relevancy, and did not mention the correctness at all, potentially due to the limited scope of tasks considered in the study.
The second reason was that participants were generally unfamiliar with a proactive assistant compared to the baseline chat, saying \textit{``I was already familiar with how to code with chatbot assistants.''}. 
Further, they already knew how to use the baseline chat well as \textit{``it feels similar to gpt-4o and copilot''} and they knew what scope of tasks the chat was helpful for (e.g., \textit{``very small and simple cases''}).
Next, we analyze how user perception translates into the actual usage of the proactive assistant.

\subsection{RQ3: How Do Participants Use Different Proactive Assistants?} \label{subsec:RQ3}

To better understand the effect of different design decisions on how participants use proactive assistants, we break down user interactions with proactive suggestions by condition.
First, we compare \texttt{Persistent Suggest} and \texttt{Suggest} conditions to explore the effect of suggestion timing.
Second, we compare \texttt{Suggest} and \texttt{Suggest and Preview} conditions to explore the effect of the level of proactivity. 
Finally, we investigate what kinds of suggestions participants tend to accept or reject.
The following discussion should be viewed as an exploratory analysis; a later addition of statistical tests via Student t-tests did not show significant differences between conditions.

\begin{figure*}[t]
\centering
\includegraphics[width=0.9\textwidth]{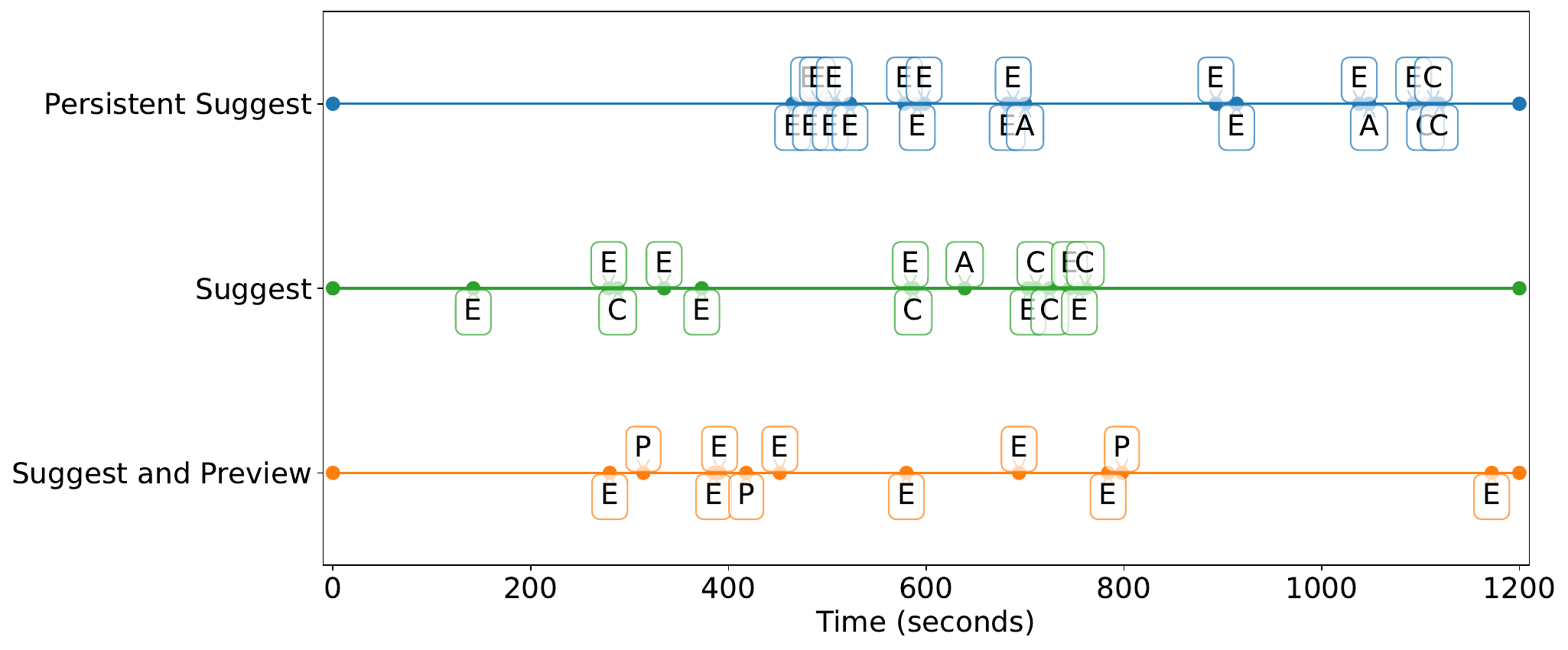}
\caption{\textbf{Comparing interactions with different proactive conditions.} Visualizing sampled participant trajectories from coding with different proactive assistants. We denote ``expanding'' a suggestion with \textbf{E}, ``accepting'' a suggestion with \textbf{A}, and ``copying'' a suggestion with \textbf{C}---all of which are available in every condition, and ``previewing'' a suggestion with \textbf{P}, which is only available in the \texttt{Suggest and Preview} condition. In the \texttt{Persistent Suggestion} condition, participants tend to expand suggestions without utilizing them further. In contrast, participants tend to copy expanded suggestions in the \texttt{Suggest} condition while participants in the \texttt{Suggest and Preview} condition tend to preview suggestions.}
\Description{\textbf{Comparing interactions with different proactive conditions.} Visualizing sampled participant trajectories from coding with different proactive assistants. We denote ``expanding'' a suggestion with \textbf{E}, ``accepting'' a suggestion with \textbf{A}, and ``copying'' a suggestion with \textbf{C}---all of which are available in every condition, and ``previewing'' a suggestion with \textbf{P}, which is only available in the \texttt{Suggest and Preview} condition. In the \texttt{Persistent Suggestion} condition, participants tend to expand suggestions without utilizing them further. In contrast, participants tend to copy expanded suggestions in the \texttt{Suggest} condition while participants in the \texttt{Suggest and Preview} condition tend to preview suggestions.}
\label{fig:telemetries}
\end{figure*}

\subsubsection{The effect of suggestion timing.} \label{subsubsec:effect_timing}
Decreasing the amount of time that the proactive assistant waits to provide suggestions means that far more suggestions were shown in the \texttt{Persistent Suggest} condition than in the \texttt{Suggest} condition.
As such, we observe that participants tended to expand more suggestions in the \texttt{Persistent Suggest} condition compared to the \texttt{Suggest} condition ($9.5 \pm 2.4$ times per task versus $6.4 \pm 2.5$ times).
However, proactive suggestions were copied more often in the \texttt{Suggest} condition compared to \texttt{Persistent Suggest} ($3.1 \pm 1.5$ as compared to $2.1\pm 0.9$ times).
Figure~\ref{fig:telemetries} shows example telemetries of participants in each condition that demonstrate this behavior.
These trends suggest that when a proactive assistant provides too many suggestions, participants can be easily distracted and thus still want to view the suggestions at a similar frequency. 
However, when the suggestions are provided too often, participants are less likely to make use of the suggestions.

\subsubsection{The effect of level of proactive assistant involvement.} \label{subsubsec:level_involve}
The ability to preview and incorporate suggestions into code led to a change in user interactions with proactive assistants.
For example, after expanding a suggestion, further interactions with suggestions were more than twice as likely to be through the preview functionality rather than copying code and integrating it into the editor by themselves ($3.1 \pm 1.5$ in the \texttt{Suggest and Preview} condition compared to $0.8 \pm 0.5$ \texttt{Suggest} condition)---an example of such an interaction pattern is shown in Figure~\ref{fig:telemetries}. 
This may also have led to an increase in manual requests for suggestions, which happened more often in the \texttt{Suggest and Preview} condition, where participants made on average $2.9\pm 0.9$ manual requests in the \texttt{Suggest and Preview} compared to an average $1.7\pm0.7$ requests in the \texttt{Suggest} condition.
These results suggest that participants may gravitate towards options that reduce their manual efforts to make changes, which include utilizing the preview functionality.

\subsubsection{What kinds of suggestions do participants tend to accept?} \label{subsub:accept}
While most proactive interactions fall under expanding, copying, or previewing suggestions, as discussed in Section~\ref{subsubsec:effect_timing} and~\ref{subsubsec:level_involve}, there were a handful of interactions where participants accepted or rejected suggestions.
We identified a total of 75 interactions where 69 were accepts and 6 rejects.
The most commonly types of accepted suggestions were those on \textit{brainstorming new functionality} and \textit{debugging (latent errors)}, with 18 occurrences each, while the least likely accepted suggestions were \textit{explaining existing code} and \textit{pointers to documentation}, with only 1 occurrence each.
These trends suggest that participants are  likely to accept more ``actionable'' suggestions, rather than ones that may be merely informative.
Interestingly, \textit{improving efficiency and modularity} suggestions were most commonly rejected suggestion types (with 5 out of 6 occurrences and the other being a \textit{completing unfinished code} suggestion). 
This behavior may have been influenced by the particular study design which emphasized code correctness rather than optimized or well-written code, suggesting that different suggestions may be preferred by programmer's with different goals.
Full results are provided in Appendix~\ref{sec:additional_results}.







\section{Discussion }

\subsection{Design Implications For Proactive Coding Assistants}\label{subsec:discussion_design}


\subsubsection{Revisiting design considerations} 
Given the findings from our experimental study, we now revisit whether
the design considerations we posed in Section~\ref{sec:design_consider} are useful guides for designing proactive coding assistants. We map qualitative feedback from participants' post-task survey responses to the five design considerations.


\textit{(1) Support efficient evaluation.} 
We observe that participants generally felt that the proactive assistant provided suggestions that had \textit{``clear answers while giving specific examples''} and allowed them to 
quickly \textit{``get a baseline understanding of the problem.''}, which provides evidence that the suggestions were presented in a way that allowed participants to efficiently understand them and determine whether to use them.
Further, the suggestions were not too distracting; another participant noted that the suggestions helped to \textit{``pinpoint specific issues while not being so obnoxious [they] couldn't ignore it to focus on other tasks.''}

\textit{(2) Support efficient utilization.}
While both the baseline non-proactive and proactive assistants can help \textit{``find errors in my code and write functions''}, a participant noted that they are \textit{``able to work faster while using the one with suggestions.'' }
In particular, many participants noted that proactive suggestions were \textit{``much easier and faster to just click on a roughly correct suggestion instead of typing out specifically what I wanted to know''} and \textit{``a lot of the suggestions can be directly pasted into the code.''}.
Further, proactive suggestions that can be directly incorporated into user code via the preview function are \textit{``immediately executable''}, and participants felt that it improved their general workflow efficiency.

\textit{(3) Show contextually relevant information.} 
Not only did participants believe that the proactive assistant often suggested relevant information (e.g., \textit{``it gave me suggestions which were related to what I wanted''}), but the assistant also often \textit{``suggested things I didn't consider and didn't know to ask for such as refactoring code and spelling errors''.}
Identifying errors seemed to be a common feature that participants liked, as \textit{``it took care of most of the errors and allowed me to focus mainly on writing my code''}.
Overall, participants felt the proactive suggestions were \textit{``more personalized''} and \textit{``more tailored''} to their needs.
In particular, a participant contrasted the proactive chat assistant with other programming assistants such as code completion tools as \textit{``[code completion] only fills the code after my cursor and don't provide insights on the already implemented code.''}
This reinforces the idea that proactive chat assistants can complement commonly used tools like GitHub Copilot.

\textit{(4) Incorporating user feedback.} 
Interestingly, this was the only design consideration that received mixed feedback from participants. 
Some participants believed \textit{``the ability to accept or reject the code it suggested was really nice''} and \textit{``intuitive''} to use while others felt like they were \textit{``unnecessary''} and \textit{``interacting with them distracts me from the task.''}
Future work may consider incorporating user feedback implicitly rather than requiring users to accept or delete suggestions. 
Related to our findings from Section~\ref{subsub:accept}, different kinds of suggestions may be preferred at different points of a programmer's workflow.

\textit{(5) Time services based on context.} 
Aside from the \texttt{Persistent Suggest} condition where participants felt  suggestions popped up too often (as discussed in Section~\ref{subsec:RQ2}), participants had positive perceptions of the proactive suggestion timings, noting that the proactive assistant was \textit{``always ready''} and comparing the proactive assistant to \textit{``a second set of eyes''} or \textit{``a second person solving [the problem] for me by my side.''}
Additionally, multiple participants suggested that the timing of suggestions can even enable them to \textit{``catch issues early''} or \textit{``fix errors early on so that they don't require refactoring the entire codebase later''}.

\subsubsection{Expanding design considerations}
We also asked participants how the proactive assistant can be improved and synthesized these responses to identify three new design considerations. 
We do not include features that were already accounted for in the different variants (e.g., participants in the \texttt{Persistent Suggest} condition wanted less frequent suggestions, and participants in the \texttt{Suggest} condition wanted to be able to incorporate suggestions into the code). These expanded design considerations present opportunities for future work to improve proactive coding assistants.

\textit{(6) Allow users to decide when they want proactive assistance.} 
Multiple participants across conditions recommended that they would like the option for when to have the proactive assistant on (e.g., \textit{``an opt-in feature''} or \textit{``could change the frequency''}), particularly because \textit{``sometimes I found it useful and sometimes I didn't''}. 
Since the goal of our study was to investigate the usage patterns and effects of a proactive assistant, providing the user the option to turn off the proactive assistant would not have been possible.
This may be a particularly necessary feature in certain contexts where the user is working on less standard coding tasks, the user may choose to turn off proactive suggestions.
Prior evaluations of LLMs have shown that models are more accurate or hallucinate less on tasks akin to those that have appeared in the training data~\citep{jain2024livecodebench,liu2024exploring}.
This feature could also be important for long-term usage and facilitating appropriate trust and reliance on AI assistance.
For example, a participant was concerned about potential over-reliance on proactive suggestions, which \textit{``can lead to a lack of understanding of a codebase. Therefore, I don't want constant suggestions.''}

\textit{(7) Incorporate additional context when generating suggestions beyond user code.} While many participants already appreciated the benefits of proactive suggestions, they had further ideas for improving the suggestion content by incorporating other aspects of user behavior.
Incorporating additional context can not only better customize suggestions to user needs, but may also be beneficial in settings where the user is working on long context tasks.
For example, participants suggested that the proactive assistant could take into account \textit{``the functions I was using'}', \textit{``what the user is focusing on, typing, or clicking,''} or \textit{``recently updated code regions.''} 
Another option suggested to incorporate further context would be to allow users to provide a natural language of their task---i.e., \textit{``if I can enter the task description somewhere as a constant part of the prompt [as] some suggestions have a slight misunderstanding of the problem''}. 

\textit{(8) Flexibly varying the amount of information shown.}
Since the current proactive assistant provides one implementation option and provides a fixed number of suggestions each time, participants noted the possibility of scaling the amount of information up and down.
For example, the assistant could provide more implementation options in the proactive suggestions, e.g., \textit{``multiple ways of solving a problem,''} particularly when different factors may matter in different contexts (e.g., runtime). 
However, there may also be times when the assistant could scale down the number of suggestions shown (e.g., \textit{``if everything was good, then tell me that everything is good.''})

\subsubsection{Comparison of findings to prior work}\label{subsec:comparison}

Finally, we discuss our findings in the context of related, prior literature on AI-assisted programming and reflect on how our results should implicate the design of future AI coding assistants. In particular, we focus on related work of two categories: findings on chat-based programming assistants and findings on other forms of proactive programming assistants.

\paragraph{Implications on chat-based programming assistants.} 
In a prior study on chat assistance, \citet{ross2023programmer} found that the ``acceptance rate'' of non-proactive chat requests, as measured by code copies, was higher than the acceptance rate of proactively generated code completions.
However, our findings suggest that proactive chat assistants can generally increase user productivity.
While these results may appear to be at odds, that is not necessarily the case as proactive suggestions can have a lower acceptance rate but still help users indirectly.
Relatedly, \citet{nam2024using} and \citet{mozannar2024realhumaneval} both highlighted issues with and the burden of appropriately prompting the chat assistant to receive helpful responses.
Our study provides evidence that a proactive assistant can reduce such a burden with proactive suggestions surfacing ideas that the user may not have thought of.
In sum, our findings suggest that chat-based assistants may benefit from incorporating \emph{some} aspects of proactivity.

\paragraph{Implications on proactive programming assistants.}
Traditional approaches to providing proactive feedback relied on collecting relevant training examples and thus were largely limited in scope in terms of the types of feedback that the system could provide~\citep{errormessage, feedback}.
In contrast, we find that with modern LLMs, our proactive assistant could provide a diverse set of suggestions that were utilized to varying degrees by participants at different points of solving the programming task.
This suggests the potential for leveraging current and even more capable models in the future as an engine for surfacing high-quality suggestions.
\citet{liveprogramming} showed previewing AI suggestions and their impacts on existing code facilitates user validation of these suggestions. 
While we do observe benefits from the \texttt{Suggest and Preview} condition, participants saw the biggest benefit from the \texttt{Suggest} condition where there was no preview functionality.
A possible reason for this may be due to the way we present previews of code edits to users in our interface, which may not be optimal for user validation and is a ripe direction for future work.

\subsection{Limitations and Future Work}

There are a few limitations to our study results and setup which we discuss here. First, our study was time-limited and consisted of tasks given to the participants rather than the tasks being self-motivated by the participants. 
The coding tasks we relied on do not span the entire set of tasks a professional programmer might encounter in their work. 
The tasks were also easily measured in the number of test cases passed, which may not be the best metric in practice.
Second, the participants in our study were entirely comprised of undergraduate and graduate students, which is not fully representative of the audience that uses AI tools for programming. 
Third, our implementation of the proactive assistant was integrated into a web-based coding environment which does not contain the full features of an enterprise IDE. 
Fourth, the conditions that we considered in our controlled lab study were limited to chat-based systems, since not all participants had experience using GitHub Copilot. 
Altogether, we encourage the reader to consider the limitations of our study design when interpreting how our findings may generalize to various real-world programming applications.
For example, while we observe that participants did not mention the correctness of suggestions, this may be an important consideration for developers who work on safety-critical code when interacting with a proactive assistant.

Our study presents multiple opportunities for future work. 
First, further work is necessary to understand the nuances of proactive assistants on a wider range of user backgrounds and coding tasks.
Additionally, future work could consider evaluating the proactive chat assistants in the presence of other programming tools like code completion to understand the interplay between different tools and their relative effects on productivity.
There are also multiple promising directions in how proactive assistants can be improved to handle longer contexts, rank suggestions, more clever ways to trigger suggestions, and reduce potential hallucinations in assistant suggestions.

\section{Conclusion}
In this paper, we explore the design and implementation of proactive AI assistants powered by large language models for chat-based programming assistants.
Our prototype considers a set of design considerations that balance the benefits of proactivity while handling its challenges.
In a randomized experimental study, we evaluated the impact of various design elements of the proactive assistant on programmer productivity and user experience.
We find that proactive assistants can generally increase the number of tasks completed and improve the user experience while coding.
Furthermore, we provide insight into how aspects of the design including the effect of suggestion timing and the level of assistant involvement change interaction patterns.
These results allow us to revisit the design implications for proactive coding assistants, validating the importance of the design considerations while introducing new considerations that allow for further customization and adaptivity to the user needs.
Altogether, this work demonstrates the value of proactivity in AI coding assistants and identifies key directions for further study to further enhance user productivity and experience.

\bibliographystyle{ACM-Reference-Format}
\bibliography{sample-base}


\begin{thebibliography}{59}


\ifx \showCODEN    \undefined \def \showCODEN     #1{\unskip}     \fi
\ifx \showDOI      \undefined \def \showDOI       #1{#1}\fi
\ifx \showISBNx    \undefined \def \showISBNx     #1{\unskip}     \fi
\ifx \showISBNxiii \undefined \def \showISBNxiii  #1{\unskip}     \fi
\ifx \showISSN     \undefined \def \showISSN      #1{\unskip}     \fi
\ifx \showLCCN     \undefined \def \showLCCN      #1{\unskip}     \fi
\ifx \shownote     \undefined \def \shownote      #1{#1}          \fi
\ifx \showarticletitle \undefined \def \showarticletitle #1{#1}   \fi
\ifx \showURL      \undefined \def \showURL       {\relax}        \fi
\providecommand\bibfield[2]{#2}
\providecommand\bibinfo[2]{#2}
\providecommand\natexlab[1]{#1}
\providecommand\showeprint[2][]{arXiv:#2}

\bibitem[Ahmed et~al\mbox{.}(2020)]%
        {ahmed2020characterizing}
\bibfield{author}{\bibinfo{person}{Umair~Z Ahmed}, \bibinfo{person}{Nisheeth Srivastava}, \bibinfo{person}{Renuka Sindhgatta}, {and} \bibinfo{person}{Amey Karkare}.} \bibinfo{year}{2020}\natexlab{}.
\newblock \showarticletitle{Characterizing the pedagogical benefits of adaptive feedback for compilation errors by novice programmers}. In \bibinfo{booktitle}{\emph{Proceedings of the ACM/IEEE 42nd International Conference on Software Engineering: Software Engineering Education and Training}}. \bibinfo{pages}{139--150}.
\newblock


\bibitem[Amershi et~al\mbox{.}(2019)]%
        {amershi2019guidelines}
\bibfield{author}{\bibinfo{person}{Saleema Amershi}, \bibinfo{person}{Dan Weld}, \bibinfo{person}{Mihaela Vorvoreanu}, \bibinfo{person}{Adam Fourney}, \bibinfo{person}{Besmira Nushi}, \bibinfo{person}{Penny Collisson}, \bibinfo{person}{Jina Suh}, \bibinfo{person}{Shamsi Iqbal}, \bibinfo{person}{Paul~N Bennett}, \bibinfo{person}{Kori Inkpen}, {et~al\mbox{.}}} \bibinfo{year}{2019}\natexlab{}.
\newblock \showarticletitle{Guidelines for human-AI interaction}. In \bibinfo{booktitle}{\emph{Proceedings of the 2019 chi conference on human factors in computing systems}}. \bibinfo{pages}{1--13}.
\newblock


\bibitem[Anthropic(2023)]%
        {claude}
\bibfield{author}{\bibinfo{person}{Anthropic}.} \bibinfo{year}{2023}\natexlab{}.
\newblock \bibinfo{title}{Meet Claude}.
\newblock
\newblock
\urldef\tempurl%
\url{https://www.anthropic.com/claude}
\showURL{%
\tempurl}


\bibitem[Avula et~al\mbox{.}(2018)]%
        {avula2018searchbots}
\bibfield{author}{\bibinfo{person}{Sandeep Avula}, \bibinfo{person}{Gordon Chadwick}, \bibinfo{person}{Jaime Arguello}, {and} \bibinfo{person}{Robert Capra}.} \bibinfo{year}{2018}\natexlab{}.
\newblock \showarticletitle{Searchbots: User engagement with chatbots during collaborative search}. In \bibinfo{booktitle}{\emph{Proceedings of the 2018 conference on human information interaction \& retrieval}}. \bibinfo{pages}{52--61}.
\newblock


\bibitem[Baraglia et~al\mbox{.}(2016)]%
        {baraglia2016initiative}
\bibfield{author}{\bibinfo{person}{Jimmy Baraglia}, \bibinfo{person}{Maya Cakmak}, \bibinfo{person}{Yukie Nagai}, \bibinfo{person}{Rajesh Rao}, {and} \bibinfo{person}{Minoru Asada}.} \bibinfo{year}{2016}\natexlab{}.
\newblock \showarticletitle{Initiative in robot assistance during collaborative task execution}. In \bibinfo{booktitle}{\emph{2016 11th ACM/IEEE international conference on human-robot interaction (HRI)}}. IEEE, \bibinfo{pages}{67--74}.
\newblock


\bibitem[Barke et~al\mbox{.}(2023)]%
        {barke2022grounded}
\bibfield{author}{\bibinfo{person}{Shraddha Barke}, \bibinfo{person}{Michael~B James}, {and} \bibinfo{person}{Nadia Polikarpova}.} \bibinfo{year}{2023}\natexlab{}.
\newblock \showarticletitle{Grounded copilot: How programmers interact with code-generating models}.
\newblock \bibinfo{journal}{\emph{Proceedings of the ACM on Programming Languages}} \bibinfo{volume}{7}, \bibinfo{number}{OOPSLA1} (\bibinfo{year}{2023}), \bibinfo{pages}{85--111}.
\newblock


\bibitem[Baym et~al\mbox{.}(2019)]%
        {baym2019intelligent}
\bibfield{author}{\bibinfo{person}{Nancy Baym}, \bibinfo{person}{Limor Shifman}, \bibinfo{person}{Christopher Persaud}, {and} \bibinfo{person}{Kelly Wagman}.} \bibinfo{year}{2019}\natexlab{}.
\newblock \showarticletitle{Intelligent failures: Clippy memes and the limits of digital assistants}.
\newblock \bibinfo{journal}{\emph{AoIR Selected Papers of Internet Research}} (\bibinfo{year}{2019}).
\newblock


\bibitem[Burnett et~al\mbox{.}(2016)]%
        {burnett2016gendermag}
\bibfield{author}{\bibinfo{person}{Margaret Burnett}, \bibinfo{person}{Simone Stumpf}, \bibinfo{person}{Jamie Macbeth}, \bibinfo{person}{Stephann Makri}, \bibinfo{person}{Laura Beckwith}, \bibinfo{person}{Irwin Kwan}, \bibinfo{person}{Anicia Peters}, {and} \bibinfo{person}{William Jernigan}.} \bibinfo{year}{2016}\natexlab{}.
\newblock \showarticletitle{GenderMag: A method for evaluating software's gender inclusiveness}.
\newblock \bibinfo{journal}{\emph{Interacting with computers}} \bibinfo{volume}{28}, \bibinfo{number}{6} (\bibinfo{year}{2016}), \bibinfo{pages}{760--787}.
\newblock


\bibitem[Chaves and Gerosa(2021)]%
        {chaves2021should}
\bibfield{author}{\bibinfo{person}{Ana~Paula Chaves} {and} \bibinfo{person}{Marco~Aurelio Gerosa}.} \bibinfo{year}{2021}\natexlab{}.
\newblock \showarticletitle{How should my chatbot interact? A survey on social characteristics in human--chatbot interaction design}.
\newblock \bibinfo{journal}{\emph{International Journal of Human--Computer Interaction}} \bibinfo{volume}{37}, \bibinfo{number}{8} (\bibinfo{year}{2021}), \bibinfo{pages}{729--758}.
\newblock


\bibitem[Chopra et~al\mbox{.}(2023)]%
        {chopra2023conversational}
\bibfield{author}{\bibinfo{person}{Bhavya Chopra}, \bibinfo{person}{Ananya Singha}, \bibinfo{person}{Anna Fariha}, \bibinfo{person}{Sumit Gulwani}, \bibinfo{person}{Chris Parnin}, \bibinfo{person}{Ashish Tiwari}, {and} \bibinfo{person}{Austin~Z Henley}.} \bibinfo{year}{2023}\natexlab{}.
\newblock \showarticletitle{Conversational Challenges in AI-Powered Data Science: Obstacles, Needs, and Design Opportunities}.
\newblock \bibinfo{journal}{\emph{arXiv preprint arXiv:2310.16164}} (\bibinfo{year}{2023}).
\newblock


\bibitem[Csikszentmihalyi and Larson(2014)]%
        {csikszentmihalyi2014flow}
\bibfield{author}{\bibinfo{person}{Mihaly Csikszentmihalyi} {and} \bibinfo{person}{Reed Larson}.} \bibinfo{year}{2014}\natexlab{}.
\newblock \bibinfo{booktitle}{\emph{Flow and the foundations of positive psychology}}. Vol.~\bibinfo{volume}{10}.
\newblock \bibinfo{publisher}{Springer}.
\newblock


\bibitem[Cui et~al\mbox{.}(2024)]%
        {cui2024productivity}
\bibfield{author}{\bibinfo{person}{Kevin~Zheyuan Cui}, \bibinfo{person}{Mert Demirer}, \bibinfo{person}{Sonia Jaffe}, \bibinfo{person}{Leon Musolff}, \bibinfo{person}{Sida Peng}, {and} \bibinfo{person}{Tobias Salz}.} \bibinfo{year}{2024}\natexlab{}.
\newblock \showarticletitle{The Productivity Effects of Generative AI: Evidence from a Field Experiment with GitHub Copilot}.
\newblock  (\bibinfo{year}{2024}).
\newblock


\bibitem[Cursor(2023)]%
        {cursor}
\bibfield{author}{\bibinfo{person}{Cursor}.} \bibinfo{year}{2023}\natexlab{}.
\newblock \bibinfo{title}{Cursor - The AI Code Editor}.
\newblock
\newblock
\urldef\tempurl%
\url{https://www.cursor.com/}
\showURL{%
\tempurl}


\bibitem[Ferdowsi et~al\mbox{.}(2024)]%
        {liveprogramming}
\bibfield{author}{\bibinfo{person}{Kasra Ferdowsi}, \bibinfo{person}{Ruanqianqian~(Lisa) Huang}, \bibinfo{person}{Michael~B. James}, \bibinfo{person}{Nadia Polikarpova}, {and} \bibinfo{person}{Sorin Lerner}.} \bibinfo{year}{2024}\natexlab{}.
\newblock \showarticletitle{Validating AI-Generated Code with Live Programming}. In \bibinfo{booktitle}{\emph{Proceedings of the 2024 CHI Conference on Human Factors in Computing Systems}} (Honolulu, HI, USA) \emph{(\bibinfo{series}{CHI '24})}. \bibinfo{publisher}{Association for Computing Machinery}, \bibinfo{address}{New York, NY, USA}, Article \bibinfo{articleno}{143}, \bibinfo{numpages}{8}~pages.
\newblock
\showISBNx{9798400703300}
\urldef\tempurl%
\url{https://doi.org/10.1145/3613904.3642495}
\showDOI{\tempurl}


\bibitem[Fitzpatrick et~al\mbox{.}(2017)]%
        {fitzpatrick2017delivering}
\bibfield{author}{\bibinfo{person}{Kathleen~Kara Fitzpatrick}, \bibinfo{person}{Alison Darcy}, {and} \bibinfo{person}{Molly Vierhile}.} \bibinfo{year}{2017}\natexlab{}.
\newblock \showarticletitle{Delivering cognitive behavior therapy to young adults with symptoms of depression and anxiety using a fully automated conversational agent (Woebot): a randomized controlled trial}.
\newblock \bibinfo{journal}{\emph{JMIR mental health}} \bibinfo{volume}{4}, \bibinfo{number}{2} (\bibinfo{year}{2017}), \bibinfo{pages}{e7785}.
\newblock


\bibitem[Github(2022)]%
        {copilot}
\bibfield{author}{\bibinfo{person}{Github}.} \bibinfo{year}{2022}\natexlab{}.
\newblock \bibinfo{title}{GitHub copilot - your AI pair programmer}.
\newblock
\newblock
\urldef\tempurl%
\url{https://github.com/features/copilot}
\showURL{%
\tempurl}


\bibitem[Gu et~al\mbox{.}(2024)]%
        {gu2023analysts}
\bibfield{author}{\bibinfo{person}{Ken Gu}, \bibinfo{person}{Ruoxi Shang}, \bibinfo{person}{Tim Althoff}, \bibinfo{person}{Chenglong Wang}, {and} \bibinfo{person}{Steven~M Drucker}.} \bibinfo{year}{2024}\natexlab{}.
\newblock \showarticletitle{How Do Analysts Understand and Verify AI-Assisted Data Analyses?}. In \bibinfo{booktitle}{\emph{Proceedings of the CHI Conference on Human Factors in Computing Systems}}. \bibinfo{pages}{1--22}.
\newblock


\bibitem[Hartmann et~al\mbox{.}(2010)]%
        {errormessage}
\bibfield{author}{\bibinfo{person}{Bj\"{o}rn Hartmann}, \bibinfo{person}{Daniel MacDougall}, \bibinfo{person}{Joel Brandt}, {and} \bibinfo{person}{Scott~R. Klemmer}.} \bibinfo{year}{2010}\natexlab{}.
\newblock \showarticletitle{What would other programmers do: suggesting solutions to error messages}. In \bibinfo{booktitle}{\emph{Proceedings of the SIGCHI Conference on Human Factors in Computing Systems}} (Atlanta, Georgia, USA) \emph{(\bibinfo{series}{CHI '10})}. \bibinfo{publisher}{Association for Computing Machinery}, \bibinfo{address}{New York, NY, USA}, \bibinfo{pages}{1019–1028}.
\newblock
\showISBNx{9781605589299}
\urldef\tempurl%
\url{https://doi.org/10.1145/1753326.1753478}
\showDOI{\tempurl}


\bibitem[Head et~al\mbox{.}(2017)]%
        {feedback}
\bibfield{author}{\bibinfo{person}{Andrew Head}, \bibinfo{person}{Elena Glassman}, \bibinfo{person}{Gustavo Soares}, \bibinfo{person}{Ryo Suzuki}, \bibinfo{person}{Lucas Figueredo}, \bibinfo{person}{Loris D'Antoni}, {and} \bibinfo{person}{Bj\"{o}rn Hartmann}.} \bibinfo{year}{2017}\natexlab{}.
\newblock \showarticletitle{Writing Reusable Code Feedback at Scale with Mixed-Initiative Program Synthesis}. In \bibinfo{booktitle}{\emph{Proceedings of the Fourth (2017) ACM Conference on Learning @ Scale}} (Cambridge, Massachusetts, USA) \emph{(\bibinfo{series}{L@S '17})}. \bibinfo{publisher}{Association for Computing Machinery}, \bibinfo{address}{New York, NY, USA}, \bibinfo{pages}{89–98}.
\newblock
\showISBNx{9781450344500}
\urldef\tempurl%
\url{https://doi.org/10.1145/3051457.3051467}
\showDOI{\tempurl}


\bibitem[Horvitz(1999)]%
        {horvitz1999principles}
\bibfield{author}{\bibinfo{person}{Eric Horvitz}.} \bibinfo{year}{1999}\natexlab{}.
\newblock \showarticletitle{Principles of mixed-initiative user interfaces}. In \bibinfo{booktitle}{\emph{Proceedings of the SIGCHI conference on Human Factors in Computing Systems}}. \bibinfo{pages}{159--166}.
\newblock


\bibitem[Hu et~al\mbox{.}(2024)]%
        {hu2024designing}
\bibfield{author}{\bibinfo{person}{Jiaxiong Hu}, \bibinfo{person}{Jingya Guo}, \bibinfo{person}{Ningjing Tang}, \bibinfo{person}{Xiaojuan Ma}, \bibinfo{person}{Yuan Yao}, \bibinfo{person}{Changyuan Yang}, {and} \bibinfo{person}{Yingqing Xu}.} \bibinfo{year}{2024}\natexlab{}.
\newblock \showarticletitle{Designing the Conversational Agent: Asking Follow-up Questions for Information Elicitation}.
\newblock \bibinfo{journal}{\emph{Proceedings of the ACM on Human-Computer Interaction}} \bibinfo{volume}{8}, \bibinfo{number}{CSCW1} (\bibinfo{year}{2024}), \bibinfo{pages}{1--30}.
\newblock


\bibitem[Jain et~al\mbox{.}(2024)]%
        {jain2024livecodebench}
\bibfield{author}{\bibinfo{person}{Naman Jain}, \bibinfo{person}{King Han}, \bibinfo{person}{Alex Gu}, \bibinfo{person}{Wen-Ding Li}, \bibinfo{person}{Fanjia Yan}, \bibinfo{person}{Tianjun Zhang}, \bibinfo{person}{Sida Wang}, \bibinfo{person}{Armando Solar-Lezama}, \bibinfo{person}{Koushik Sen}, {and} \bibinfo{person}{Ion Stoica}.} \bibinfo{year}{2024}\natexlab{}.
\newblock \showarticletitle{Livecodebench: Holistic and contamination free evaluation of large language models for code}.
\newblock \bibinfo{journal}{\emph{arXiv preprint arXiv:2403.07974}} (\bibinfo{year}{2024}).
\newblock


\bibitem[Kazemitabaar et~al\mbox{.}(2023)]%
        {kazemitabaar2023studying}
\bibfield{author}{\bibinfo{person}{Majeed Kazemitabaar}, \bibinfo{person}{Justin Chow}, \bibinfo{person}{Carl Ka~To Ma}, \bibinfo{person}{Barbara~J Ericson}, \bibinfo{person}{David Weintrop}, {and} \bibinfo{person}{Tovi Grossman}.} \bibinfo{year}{2023}\natexlab{}.
\newblock \showarticletitle{Studying the effect of ai code generators on supporting novice learners in introductory programming}. In \bibinfo{booktitle}{\emph{Proceedings of the 2023 CHI Conference on Human Factors in Computing Systems}}. \bibinfo{pages}{1--23}.
\newblock


\bibitem[Kazemitabaar et~al\mbox{.}(2024)]%
        {kazemitabaar2024codeaid}
\bibfield{author}{\bibinfo{person}{Majeed Kazemitabaar}, \bibinfo{person}{Runlong Ye}, \bibinfo{person}{Xiaoning Wang}, \bibinfo{person}{Austin~Z Henley}, \bibinfo{person}{Paul Denny}, \bibinfo{person}{Michelle Craig}, {and} \bibinfo{person}{Tovi Grossman}.} \bibinfo{year}{2024}\natexlab{}.
\newblock \showarticletitle{CodeAid: Evaluating a Classroom Deployment of an LLM-based Programming Assistant that Balances Student and Educator Needs}.
\newblock  (\bibinfo{year}{2024}).
\newblock


\bibitem[Lee et~al\mbox{.}(2024)]%
        {lee2024design}
\bibfield{author}{\bibinfo{person}{Mina Lee}, \bibinfo{person}{Katy~Ilonka Gero}, \bibinfo{person}{John Joon~Young Chung}, \bibinfo{person}{Simon~Buckingham Shum}, \bibinfo{person}{Vipul Raheja}, \bibinfo{person}{Hua Shen}, \bibinfo{person}{Subhashini Venugopalan}, \bibinfo{person}{Thiemo Wambsganss}, \bibinfo{person}{David Zhou}, \bibinfo{person}{Emad~A Alghamdi}, {et~al\mbox{.}}} \bibinfo{year}{2024}\natexlab{}.
\newblock \showarticletitle{A Design Space for Intelligent and Interactive Writing Assistants}. In \bibinfo{booktitle}{\emph{Proceedings of the CHI Conference on Human Factors in Computing Systems}}. \bibinfo{pages}{1--35}.
\newblock


\bibitem[Lee et~al\mbox{.}(2022)]%
        {lee2022coauthor}
\bibfield{author}{\bibinfo{person}{Mina Lee}, \bibinfo{person}{Percy Liang}, {and} \bibinfo{person}{Qian Yang}.} \bibinfo{year}{2022}\natexlab{}.
\newblock \showarticletitle{Coauthor: Designing a human-ai collaborative writing dataset for exploring language model capabilities}. In \bibinfo{booktitle}{\emph{CHI Conference on Human Factors in Computing Systems}}. \bibinfo{pages}{1--19}.
\newblock


\bibitem[Lerner(2020)]%
        {liveprogramming1}
\bibfield{author}{\bibinfo{person}{Sorin Lerner}.} \bibinfo{year}{2020}\natexlab{}.
\newblock \showarticletitle{Projection Boxes: On-the-fly Reconfigurable Visualization for Live Programming}. In \bibinfo{booktitle}{\emph{Proceedings of the 2020 CHI Conference on Human Factors in Computing Systems}} (Honolulu, HI, USA) \emph{(\bibinfo{series}{CHI '20})}. \bibinfo{publisher}{Association for Computing Machinery}, \bibinfo{address}{New York, NY, USA}, \bibinfo{pages}{1–7}.
\newblock
\showISBNx{9781450367080}
\urldef\tempurl%
\url{https://doi.org/10.1145/3313831.3376494}
\showDOI{\tempurl}


\bibitem[Liang et~al\mbox{.}(2023)]%
        {liang2023large}
\bibfield{author}{\bibinfo{person}{Jenny~T Liang}, \bibinfo{person}{Chenyang Yang}, {and} \bibinfo{person}{Brad~A Myers}.} \bibinfo{year}{2023}\natexlab{}.
\newblock \showarticletitle{A large-scale survey on the usability of ai programming assistants: Successes and challenges}. In \bibinfo{booktitle}{\emph{2024 IEEE/ACM 46th International Conference on Software Engineering (ICSE)}}. IEEE Computer Society, \bibinfo{pages}{605--617}.
\newblock


\bibitem[Liao et~al\mbox{.}(2016)]%
        {liao2016can}
\bibfield{author}{\bibinfo{person}{Q~Vera Liao}, \bibinfo{person}{Matthew Davis}, \bibinfo{person}{Werner Geyer}, \bibinfo{person}{Michael Muller}, {and} \bibinfo{person}{N~Sadat Shami}.} \bibinfo{year}{2016}\natexlab{}.
\newblock \showarticletitle{What can you do? Studying social-agent orientation and agent proactive interactions with an agent for employees}. In \bibinfo{booktitle}{\emph{Proceedings of the 2016 acm conference on designing interactive systems}}. \bibinfo{pages}{264--275}.
\newblock


\bibitem[Lin et~al\mbox{.}(2022)]%
        {lin2022inferring}
\bibfield{author}{\bibinfo{person}{Jessy Lin}, \bibinfo{person}{Daniel Fried}, \bibinfo{person}{Dan Klein}, {and} \bibinfo{person}{Anca Dragan}.} \bibinfo{year}{2022}\natexlab{}.
\newblock \showarticletitle{Inferring Rewards from Language in Context}. In \bibinfo{booktitle}{\emph{Proceedings of the 60th Annual Meeting of the Association for Computational Linguistics (Volume 1: Long Papers)}}. \bibinfo{pages}{8546--8560}.
\newblock


\bibitem[Liu et~al\mbox{.}(2024a)]%
        {liu2024exploring}
\bibfield{author}{\bibinfo{person}{Fang Liu}, \bibinfo{person}{Yang Liu}, \bibinfo{person}{Lin Shi}, \bibinfo{person}{Houkun Huang}, \bibinfo{person}{Ruifeng Wang}, \bibinfo{person}{Zhen Yang}, \bibinfo{person}{Li Zhang}, \bibinfo{person}{Zhongqi Li}, {and} \bibinfo{person}{Yuchi Ma}.} \bibinfo{year}{2024}\natexlab{a}.
\newblock \showarticletitle{Exploring and evaluating hallucinations in llm-powered code generation}.
\newblock \bibinfo{journal}{\emph{arXiv preprint arXiv:2404.00971}} (\bibinfo{year}{2024}).
\newblock


\bibitem[Liu et~al\mbox{.}(2024b)]%
        {liu2024compeer}
\bibfield{author}{\bibinfo{person}{Tianjian Liu}, \bibinfo{person}{Hongzheng Zhao}, \bibinfo{person}{Yuheng Liu}, \bibinfo{person}{Xingbo Wang}, {and} \bibinfo{person}{Zhenhui Peng}.} \bibinfo{year}{2024}\natexlab{b}.
\newblock \showarticletitle{ComPeer: A Generative Conversational Agent for Proactive Peer Support}.
\newblock \bibinfo{journal}{\emph{arXiv preprint arXiv:2407.18064}} (\bibinfo{year}{2024}).
\newblock


\bibitem[Luger and Sellen(2016)]%
        {luger2016like}
\bibfield{author}{\bibinfo{person}{Ewa Luger} {and} \bibinfo{person}{Abigail Sellen}.} \bibinfo{year}{2016}\natexlab{}.
\newblock \showarticletitle{" Like Having a Really Bad PA" The Gulf between User Expectation and Experience of Conversational Agents}. In \bibinfo{booktitle}{\emph{Proceedings of the 2016 CHI conference on human factors in computing systems}}. \bibinfo{pages}{5286--5297}.
\newblock


\bibitem[Meurisch et~al\mbox{.}(2020)]%
        {meurisch2020exploring}
\bibfield{author}{\bibinfo{person}{Christian Meurisch}, \bibinfo{person}{Cristina~A Mihale-Wilson}, \bibinfo{person}{Adrian Hawlitschek}, \bibinfo{person}{Florian Giger}, \bibinfo{person}{Florian M{\"u}ller}, \bibinfo{person}{Oliver Hinz}, {and} \bibinfo{person}{Max M{\"u}hlh{\"a}user}.} \bibinfo{year}{2020}\natexlab{}.
\newblock \showarticletitle{Exploring user expectations of proactive AI systems}.
\newblock \bibinfo{journal}{\emph{Proceedings of the ACM on Interactive, Mobile, Wearable and Ubiquitous Technologies}} \bibinfo{volume}{4}, \bibinfo{number}{4} (\bibinfo{year}{2020}), \bibinfo{pages}{1--22}.
\newblock


\bibitem[Mozannar et~al\mbox{.}(2024a)]%
        {mozannar2022reading}
\bibfield{author}{\bibinfo{person}{Hussein Mozannar}, \bibinfo{person}{Gagan Bansal}, \bibinfo{person}{Adam Fourney}, {and} \bibinfo{person}{Eric Horvitz}.} \bibinfo{year}{2024}\natexlab{a}.
\newblock \showarticletitle{Reading between the lines: Modeling user behavior and costs in AI-assisted programming}. In \bibinfo{booktitle}{\emph{Proceedings of the CHI Conference on Human Factors in Computing Systems}}. \bibinfo{pages}{1--16}.
\newblock


\bibitem[Mozannar et~al\mbox{.}(2024b)]%
        {mozannar2024show}
\bibfield{author}{\bibinfo{person}{Hussein Mozannar}, \bibinfo{person}{Gagan Bansal}, \bibinfo{person}{Adam Fourney}, {and} \bibinfo{person}{Eric Horvitz}.} \bibinfo{year}{2024}\natexlab{b}.
\newblock \showarticletitle{When to show a suggestion? Integrating human feedback in AI-assisted programming}. In \bibinfo{booktitle}{\emph{Proceedings of the AAAI Conference on Artificial Intelligence}}, Vol.~\bibinfo{volume}{38}. \bibinfo{pages}{10137--10144}.
\newblock


\bibitem[Mozannar et~al\mbox{.}(2024c)]%
        {mozannar2024realhumaneval}
\bibfield{author}{\bibinfo{person}{Hussein Mozannar}, \bibinfo{person}{Valerie Chen}, \bibinfo{person}{Mohammed Alsobay}, \bibinfo{person}{Subhro Das}, \bibinfo{person}{Sebastian Zhao}, \bibinfo{person}{Dennis Wei}, \bibinfo{person}{Manish Nagireddy}, \bibinfo{person}{Prasanna Sattigeri}, \bibinfo{person}{Ameet Talwalkar}, {and} \bibinfo{person}{David Sontag}.} \bibinfo{year}{2024}\natexlab{c}.
\newblock \showarticletitle{The RealHumanEval: Evaluating Large Language Models' Abilities to Support Programmers}.
\newblock \bibinfo{journal}{\emph{arXiv preprint arXiv:2404.02806}} (\bibinfo{year}{2024}).
\newblock


\bibitem[Nam et~al\mbox{.}(2024)]%
        {nam2024using}
\bibfield{author}{\bibinfo{person}{Daye Nam}, \bibinfo{person}{Andrew Macvean}, \bibinfo{person}{Vincent Hellendoorn}, \bibinfo{person}{Bogdan Vasilescu}, {and} \bibinfo{person}{Brad Myers}.} \bibinfo{year}{2024}\natexlab{}.
\newblock \showarticletitle{Using an llm to help with code understanding}. In \bibinfo{booktitle}{\emph{Proceedings of the IEEE/ACM 46th International Conference on Software Engineering}}. \bibinfo{pages}{1--13}.
\newblock


\bibitem[OpenAI(2022)]%
        {chatgpt}
\bibfield{author}{\bibinfo{person}{OpenAI}.} \bibinfo{year}{2022}\natexlab{}.
\newblock \bibinfo{title}{ChatGPT: Optimizing Language Models for Dialogue}.
\newblock
\newblock
\urldef\tempurl%
\url{https://openai.com/blog/chatgpt/}
\showURL{%
\tempurl}


\bibitem[Park et~al\mbox{.}(2023)]%
        {park2023generative}
\bibfield{author}{\bibinfo{person}{Joon~Sung Park}, \bibinfo{person}{Joseph~C O'Brien}, \bibinfo{person}{Carrie~J Cai}, \bibinfo{person}{Meredith~Ringel Morris}, \bibinfo{person}{Percy Liang}, {and} \bibinfo{person}{Michael~S Bernstein}.} \bibinfo{year}{2023}\natexlab{}.
\newblock \showarticletitle{Generative agents: Interactive simulacra of human behavior}.
\newblock \bibinfo{journal}{\emph{arXiv preprint arXiv:2304.03442}} (\bibinfo{year}{2023}).
\newblock


\bibitem[Peng et~al\mbox{.}(2023)]%
        {peng2023impact}
\bibfield{author}{\bibinfo{person}{Sida Peng}, \bibinfo{person}{Eirini Kalliamvakou}, \bibinfo{person}{Peter Cihon}, {and} \bibinfo{person}{Mert Demirer}.} \bibinfo{year}{2023}\natexlab{}.
\newblock \showarticletitle{The impact of ai on developer productivity: Evidence from github copilot}.
\newblock \bibinfo{journal}{\emph{arXiv preprint arXiv:2302.06590}} (\bibinfo{year}{2023}).
\newblock


\bibitem[Peng et~al\mbox{.}(2019)]%
        {peng2019design}
\bibfield{author}{\bibinfo{person}{Zhenhui Peng}, \bibinfo{person}{Yunhwan Kwon}, \bibinfo{person}{Jiaan Lu}, \bibinfo{person}{Ziming Wu}, {and} \bibinfo{person}{Xiaojuan Ma}.} \bibinfo{year}{2019}\natexlab{}.
\newblock \showarticletitle{Design and evaluation of service robot's proactivity in decision-making support process}. In \bibinfo{booktitle}{\emph{Proceedings of the 2019 CHI Conference on Human Factors in Computing Systems}}. \bibinfo{pages}{1--13}.
\newblock


\bibitem[Prather et~al\mbox{.}(2023)]%
        {prather2023its}
\bibfield{author}{\bibinfo{person}{James Prather}, \bibinfo{person}{Brent~N. Reeves}, \bibinfo{person}{Paul Denny}, \bibinfo{person}{Brett~A. Becker}, \bibinfo{person}{Juho Leinonen}, \bibinfo{person}{Andrew Luxton-Reilly}, \bibinfo{person}{Garrett Powell}, \bibinfo{person}{James Finnie-Ansley}, {and} \bibinfo{person}{Eddie~Antonio Santos}.} \bibinfo{year}{2023}\natexlab{}.
\newblock \showarticletitle{“It’s Weird That It Knows What I Want”: Usability and Interactions with Copilot for Novice Programmers}.
\newblock \bibinfo{journal}{\emph{ACM Trans. Comput.-Hum. Interact.}} \bibinfo{volume}{31}, \bibinfo{number}{1}, Article \bibinfo{articleno}{4} (\bibinfo{date}{nov} \bibinfo{year}{2023}), \bibinfo{numpages}{31}~pages.
\newblock
\showISSN{1073-0516}
\urldef\tempurl%
\url{https://doi.org/10.1145/3617367}
\showDOI{\tempurl}


\bibitem[Ross et~al\mbox{.}(2023)]%
        {ross2023programmer}
\bibfield{author}{\bibinfo{person}{Steven~I Ross}, \bibinfo{person}{Fernando Martinez}, \bibinfo{person}{Stephanie Houde}, \bibinfo{person}{Michael Muller}, {and} \bibinfo{person}{Justin~D Weisz}.} \bibinfo{year}{2023}\natexlab{}.
\newblock \showarticletitle{The programmer’s assistant: Conversational interaction with a large language model for software development}. In \bibinfo{booktitle}{\emph{Proceedings of the 28th International Conference on Intelligent User Interfaces}}. \bibinfo{pages}{491--514}.
\newblock


\bibitem[Shaer et~al\mbox{.}(2024)]%
        {shaer2024ai}
\bibfield{author}{\bibinfo{person}{Orit Shaer}, \bibinfo{person}{Angelora Cooper}, \bibinfo{person}{Osnat Mokryn}, \bibinfo{person}{Andrew~L Kun}, {and} \bibinfo{person}{Hagit Ben~Shoshan}.} \bibinfo{year}{2024}\natexlab{}.
\newblock \showarticletitle{AI-Augmented Brainwriting: Investigating the use of LLMs in group ideation}. In \bibinfo{booktitle}{\emph{Proceedings of the CHI Conference on Human Factors in Computing Systems}}. \bibinfo{pages}{1--17}.
\newblock


\bibitem[Silvervarg and J{\"o}nsson(2013)]%
        {silvervarg2013iterative}
\bibfield{author}{\bibinfo{person}{Annika Silvervarg} {and} \bibinfo{person}{Arne J{\"o}nsson}.} \bibinfo{year}{2013}\natexlab{}.
\newblock \showarticletitle{Iterative development and evaluation of a social conversational agent}. In \bibinfo{booktitle}{\emph{Proceedings of the Sixth International Joint Conference on Natural Language Processing}}. \bibinfo{pages}{1223--1229}.
\newblock


\bibitem[Tallyn et~al\mbox{.}(2018)]%
        {tallyn2018ethnobot}
\bibfield{author}{\bibinfo{person}{Ella Tallyn}, \bibinfo{person}{Hector Fried}, \bibinfo{person}{Rory Gianni}, \bibinfo{person}{Amy Isard}, {and} \bibinfo{person}{Chris Speed}.} \bibinfo{year}{2018}\natexlab{}.
\newblock \showarticletitle{The ethnobot: Gathering ethnographies in the age of IoT}. In \bibinfo{booktitle}{\emph{Proceedings of the 2018 CHI conference on human factors in computing systems}}. \bibinfo{pages}{1--13}.
\newblock


\bibitem[Toxtli et~al\mbox{.}(2018)]%
        {toxtli2018understanding}
\bibfield{author}{\bibinfo{person}{Carlos Toxtli}, \bibinfo{person}{Andr{\'e}s Monroy-Hern{\'a}ndez}, {and} \bibinfo{person}{Justin Cranshaw}.} \bibinfo{year}{2018}\natexlab{}.
\newblock \showarticletitle{Understanding chatbot-mediated task management}. In \bibinfo{booktitle}{\emph{Proceedings of the 2018 CHI conference on human factors in computing systems}}. \bibinfo{pages}{1--6}.
\newblock


\bibitem[Vaithilingam et~al\mbox{.}(2022)]%
        {vaithilingam2022expectation}
\bibfield{author}{\bibinfo{person}{Priyan Vaithilingam}, \bibinfo{person}{Tianyi Zhang}, {and} \bibinfo{person}{Elena~L Glassman}.} \bibinfo{year}{2022}\natexlab{}.
\newblock \showarticletitle{Expectation vs. Experience: Evaluating the Usability of Code Generation Tools Powered by Large Language Models}. In \bibinfo{booktitle}{\emph{CHI Conference on Human Factors in Computing Systems Extended Abstracts}}. \bibinfo{pages}{1--7}.
\newblock


\bibitem[Vasconcelos et~al\mbox{.}(2023)]%
        {vasconcelos2023generation}
\bibfield{author}{\bibinfo{person}{Helena Vasconcelos}, \bibinfo{person}{Gagan Bansal}, \bibinfo{person}{Adam Fourney}, \bibinfo{person}{Q~Vera Liao}, {and} \bibinfo{person}{Jennifer~Wortman Vaughan}.} \bibinfo{year}{2023}\natexlab{}.
\newblock \showarticletitle{Generation probabilities are not enough: Exploring the effectiveness of uncertainty highlighting in AI-powered code completions}.
\newblock \bibinfo{journal}{\emph{arXiv preprint arXiv:2302.07248}} (\bibinfo{year}{2023}).
\newblock


\bibitem[Winkler and Roos(2019)]%
        {winkler2019bringing}
\bibfield{author}{\bibinfo{person}{Rainer Winkler} {and} \bibinfo{person}{Julian Roos}.} \bibinfo{year}{2019}\natexlab{}.
\newblock \showarticletitle{Bringing AI into the classroom: Designing smart personal assistants as learning tutors}.
\newblock  (\bibinfo{year}{2019}).
\newblock


\bibitem[Xiao et~al\mbox{.}(2023)]%
        {xiao2023devgpt}
\bibfield{author}{\bibinfo{person}{Tao Xiao}, \bibinfo{person}{Christoph Treude}, \bibinfo{person}{Hideaki Hata}, {and} \bibinfo{person}{Kenichi Matsumoto}.} \bibinfo{year}{2023}\natexlab{}.
\newblock \showarticletitle{DevGPT: Studying Developer-ChatGPT Conversations}.
\newblock \bibinfo{journal}{\emph{arXiv preprint arXiv:2309.03914}} (\bibinfo{year}{2023}).
\newblock


\bibitem[Yan et~al\mbox{.}(2024)]%
        {yan2024ivie}
\bibfield{author}{\bibinfo{person}{Litao Yan}, \bibinfo{person}{Alyssa Hwang}, \bibinfo{person}{Zhiyuan Wu}, {and} \bibinfo{person}{Andrew Head}.} \bibinfo{year}{2024}\natexlab{}.
\newblock \showarticletitle{Ivie: Lightweight anchored explanations of just-generated code}. In \bibinfo{booktitle}{\emph{Proceedings of the CHI Conference on Human Factors in Computing Systems}}. \bibinfo{pages}{1--15}.
\newblock


\bibitem[Yang et~al\mbox{.}(2024)]%
        {yang2024swe}
\bibfield{author}{\bibinfo{person}{John Yang}, \bibinfo{person}{Carlos~E Jimenez}, \bibinfo{person}{Alexander Wettig}, \bibinfo{person}{Kilian Lieret}, \bibinfo{person}{Shunyu Yao}, \bibinfo{person}{Karthik Narasimhan}, {and} \bibinfo{person}{Ofir Press}.} \bibinfo{year}{2024}\natexlab{}.
\newblock \showarticletitle{Swe-agent: Agent-computer interfaces enable automated software engineering}.
\newblock \bibinfo{journal}{\emph{arXiv preprint arXiv:2405.15793}} (\bibinfo{year}{2024}).
\newblock


\bibitem[Yao et~al\mbox{.}(2024)]%
        {yao2024tau}
\bibfield{author}{\bibinfo{person}{Shunyu Yao}, \bibinfo{person}{Noah Shinn}, \bibinfo{person}{Pedram Razavi}, {and} \bibinfo{person}{Karthik Narasimhan}.} \bibinfo{year}{2024}\natexlab{}.
\newblock \showarticletitle{$\tau$-bench: A Benchmark for Tool-Agent-User Interaction in Real-World Domains}.
\newblock \bibinfo{journal}{\emph{arXiv preprint arXiv:2406.12045}} (\bibinfo{year}{2024}).
\newblock


\bibitem[Zhang et~al\mbox{.}(2015)]%
        {zhang2015human}
\bibfield{author}{\bibinfo{person}{Yu Zhang}, \bibinfo{person}{Vignesh Narayanan}, \bibinfo{person}{Tathagata Chakraborti}, {and} \bibinfo{person}{Subbarao Kambhampati}.} \bibinfo{year}{2015}\natexlab{}.
\newblock \showarticletitle{A human factors analysis of proactive support in human-robot teaming}. In \bibinfo{booktitle}{\emph{2015 IEEE/RSJ International Conference on Intelligent Robots and Systems (IROS)}}. IEEE, \bibinfo{pages}{3586--3593}.
\newblock


\bibitem[Zhao et~al\mbox{.}(2024)]%
        {zhao2024wildchat}
\bibfield{author}{\bibinfo{person}{Wenting Zhao}, \bibinfo{person}{Xiang Ren}, \bibinfo{person}{Jack Hessel}, \bibinfo{person}{Claire Cardie}, \bibinfo{person}{Yejin Choi}, {and} \bibinfo{person}{Yuntian Deng}.} \bibinfo{year}{2024}\natexlab{}.
\newblock \showarticletitle{WildChat: 1M Chat{GPT} Interaction Logs in the Wild}. In \bibinfo{booktitle}{\emph{The Twelfth International Conference on Learning Representations}}.
\newblock
\urldef\tempurl%
\url{https://openreview.net/forum?id=Bl8u7ZRlbM}
\showURL{%
\tempurl}


\bibitem[Zheng et~al\mbox{.}(2023)]%
        {zheng2023lmsyschat1m}
\bibfield{author}{\bibinfo{person}{Lianmin Zheng}, \bibinfo{person}{Wei-Lin Chiang}, \bibinfo{person}{Ying Sheng}, \bibinfo{person}{Tianle Li}, \bibinfo{person}{Siyuan Zhuang}, \bibinfo{person}{Zhanghao Wu}, \bibinfo{person}{Yonghao Zhuang}, \bibinfo{person}{Zhuohan Li}, \bibinfo{person}{Zi Lin}, \bibinfo{person}{Eric.~P Xing}, \bibinfo{person}{Joseph~E. Gonzalez}, \bibinfo{person}{Ion Stoica}, {and} \bibinfo{person}{Hao Zhang}.} \bibinfo{year}{2023}\natexlab{}.
\newblock \bibinfo{title}{LMSYS-Chat-1M: A Large-Scale Real-World LLM Conversation Dataset}.
\newblock
\newblock
\showeprint[arxiv]{2309.11998}~[cs.CL]


\bibitem[Zhou et~al\mbox{.}(2023)]%
        {zhou2023webarena}
\bibfield{author}{\bibinfo{person}{Shuyan Zhou}, \bibinfo{person}{Frank~F Xu}, \bibinfo{person}{Hao Zhu}, \bibinfo{person}{Xuhui Zhou}, \bibinfo{person}{Robert Lo}, \bibinfo{person}{Abishek Sridhar}, \bibinfo{person}{Xianyi Cheng}, \bibinfo{person}{Tianyue Ou}, \bibinfo{person}{Yonatan Bisk}, \bibinfo{person}{Daniel Fried}, {et~al\mbox{.}}} \bibinfo{year}{2023}\natexlab{}.
\newblock \showarticletitle{Webarena: A realistic web environment for building autonomous agents}.
\newblock \bibinfo{journal}{\emph{arXiv preprint arXiv:2307.13854}} (\bibinfo{year}{2023}).
\newblock


\end{thebibliography}

\clearpage
\appendix
\section{User Study Details}

\subsection{Instructions}

\paragraph{General instructions} Welcome to python coding study!
\begin{itemize}
    \item You will be writing code in Python only, and use only standard python libraries in addition to numpy.
    \item You will have 40 minutes total where you will try to solve as many coding tasks as possible one at a time.
    \item It is NOT allowed to use any outside resources to solve the coding questions (e.g. Google, StackOverflow, ChatGPT), your compensation is tied to effort only.
\end{itemize}

\paragraph{Baseline condition}

You will write code in the interface above: a code editor equipped with an AI assistant chatbot. The chatbot does not have access to either the code editor or the task description.
\begin{itemize}
    \item Please write python code only in the editor. We only support standard python3.8 packages and numpy.
    \item You can run your code by pressing the "Run" button and the output will be in the output box at the bottom in grey.
    \item Press the ``Submit'' button to submit your code. You can only submit your code once for each task.
    \item You are provided with a chat interface that you can use to help you with the coding tasks.
    \item Pressing the "Clear" button on the top of the chat window will clear the current chat history. Chat history will be automatically cleared when you move to the next task.
    \item If the chat output contains a code snippet, you can click the icon to copy the code to your clipboard.
    Please be aware that its output is not always correct.
\end{itemize}

\paragraph{Proactive condition}

You have access to the Proactive Coding Assistant. In addition to the regular chatbot functionality, the assistant will occasionally provide suggestions when you are stuck! Please try to use it in the study if the suggestions seem helpful. Here’s how:

Click on a suggestion title for the following options:

\begin{itemize}
    \item Previewing: Click "Preview" to see how the proactive assistant incorporates the suggestion into your code, which can then accept or hide the changes. Use the copy icon to copy the code to your clipboard.
    \item Accepting: click “Accept” to add the expanded suggestion to the chat window and ask follow-up questions.
    \item Deleting: click “Clear all” to remove all current suggestions. Use “Delete” to remove a single expanded suggestion.
\end{itemize}

Other features:
\begin{itemize}
    \item If a regular chat message contains code, you can also have the proactive assistant help you incorporate into your code.
    \item You can request suggestions by clicking on the “Suggest” button at the top of the chat window, or using the shortcut [Ctrl + Enter] (Windows) or [Cmd + Enter] (Mac) in the code editor.
    \item After running your code, the assistant analyzes the output or error messages and offers debugging suggestions. You can interact with these suggestions in the same way.
\end{itemize}

Warning: the assistant does not have access to the task description, but it has access to your code editor and can provide context-aware suggestions based on your code and cursor position. Suggestions are not always perfect, and the code provided may be inaccurate. In addition the suggestions may use packages that we do not support (we only support numpy). Always review suggestions thoroughly before integrating them into your code.

You can review this tutorial again by clicking on the “Show Instructions” button at the top of the page, then clicking "Next".

\subsection{Post-study Questionnaire}

\begin{itemize}
    \item Overall, how useful was the non-proactive chat assistant (e.g., the one that did not automatically provide suggestions)? (on a scale of 1-10, where 1 is least helpful and 10 is most helpful)
    \item Overall, how useful was the proactive chat assistant (e.g., the one that automatically provided suggestions)? (on a scale of 1-10, where 1 is least helpful and 10 is most helpful)
    \item Briefly explain the reasoning for your ratings above. (free response)
    \item What kind of suggestions did you find most helpful? (free response)
    \item In what ways could the proactive chat assistant suggestions be improved? 
    \item In what ways could the proactive chat assistant interface be improved?
\end{itemize}

\section{Task Design}

\subsection{Tasks} We consider a total of 4 tasks in our study, where each task comprises multiple sub-tasks.

\paragraph{Task 1: Storefront}

\begin{quote}
You are a freelance software engineer hired to design the backend of an online store.
To get started, we have provided starter code for the Store class. 

Your goal is to add more functionality to the online store to make it more complete. Please complete Sub-Task 1 first, then you can complete Sub-Tasks 2 and 3 in any order.

Sub-Task 1: major additions
- write the Order and Product class, which should work with the current code in the Store class

Sub-Task 2: add the following functionality to the Store class
- write function `apply\_discount\_to\_order(self, order\_id, discount)` 
- write function `check\_order\_status(self, order\_id)` 

Sub-Task 3: create one additional feature that you think might be good to have

Make sure to write test cases to demonstrate that the online store works as intended.

You can only submit your code once for this task. Please only submit your code after you have completed as many sub-tasks as you can.
\end{quote}

\paragraph{Task 2: To-do List}

\begin{quote}
    You are an application developer working on the backend of a to-do list app.
You are working with some existing code your colleague left off. 
Each task takes in a `description` (string), `category` (string), `priority` (string, `"High"`, `"Medium"`, or `"Low"`), and optional `date\_due` (datetime object).

Your goal is to add more functionality to the to-do list to make it more complete. You can complete the subtasks in any order.

Sub-Task 1: minor fixes
- when a task is overdue, change the `\_\_str\_\_` function of the task to reflect the task is overdue by adding `(OVERDUE)` after the due date
- when adding a new task, do not allow the user to add it if the due date is in the past
- there seems to be a bug in the code, please find and fix it
- the efficiency of editing a task can be improved, please edit this functionality

Sub-Task 2: add features
- modify the `list\_all(self)` function by adding an argument `show\_completed` which prints only unfinished tasks when set to False
- add a function `list\_by\_priority(self)` to ToDoList which prints the task in order of priority, from high to low

Make sure to write test cases to demonstrate that the to-do list works as intended.

You can only submit your code once for this task. Please only submit your code after you have completed as many sub-tasks as you can.
\end{quote}

\paragraph{Task 3: Sales analysis}

\begin{quote}
    You are a data analyst working for a global retailer.
You are working with product sales data from the past year. 

The sales data is a 4-dimensional numpy array with shape (3, 12, 10, 100) and the following structure:

- Axis 0: Different regions

- Axis 1: Different months

- Axis 2: Different stores

- Axis 3: Product sales figures

Given the described sales data, you need to use ONLY numpy packages to complete the following sub-tasks:

1. write function `total\_sales\_per\_region(data)` that calculates total sales number for each region

2. write function `cumulative\_sales(data)` that computes the cumulative sales over time for each product across all regions and stores

3. write function `top\_products\_by\_sales(data, k)` that computes top k best selling product id for each month across all regions and stores

4. write function `temporal\_correlation(data)` that calculates pairwise temporal correlations over time for each product

Your goal is to complete the above sub-tasks. Make sure to write test cases to demonstrate that the code works as intended.

You can only submit your code once for this task. Please only submit your code after you have completed as many sub-tasks as you can.
\end{quote}

\paragraph{Task 4: Weather trends}

\begin{quote}
    You are a data analyst working for a meteorological department. 
Your task is to analyze temperature data for a given month to gain insights into weather patterns and trends. 

Given temperature data for a month, you need to use ONLY numpy packages to complete the following sub-tasks:

1. fill out `classify\_temps(data)` which classifies each day's temperature into categories such as 'Freezing', 'Moderate', and 'Hot'.

2. write function `clip\_temps(data)` which clips any extreme temperature values to ensure they fall within -10 and 40.

3. write function `compute\_moving\_avg(data, window\_size)` which calculates the moving average of temperatures over a specified window\_size (e.g., 7 days).

4. write function `compute\_weekly\_avg(data)` which calculates weekly average temperatures.

Your goal is to complete the above sub-tasks. The provided code is incomplete and may not be fully correct. 

Make sure to write test cases to demonstrate that the code works as intended.

You can only submit your code once for this task. Please only submit your code after you have completed as many sub-tasks as you can.
\end{quote}

\section{Additional results}\label{sec:additional_results}

\begin{figure*}[h]
\centering
\includegraphics[width=\textwidth]{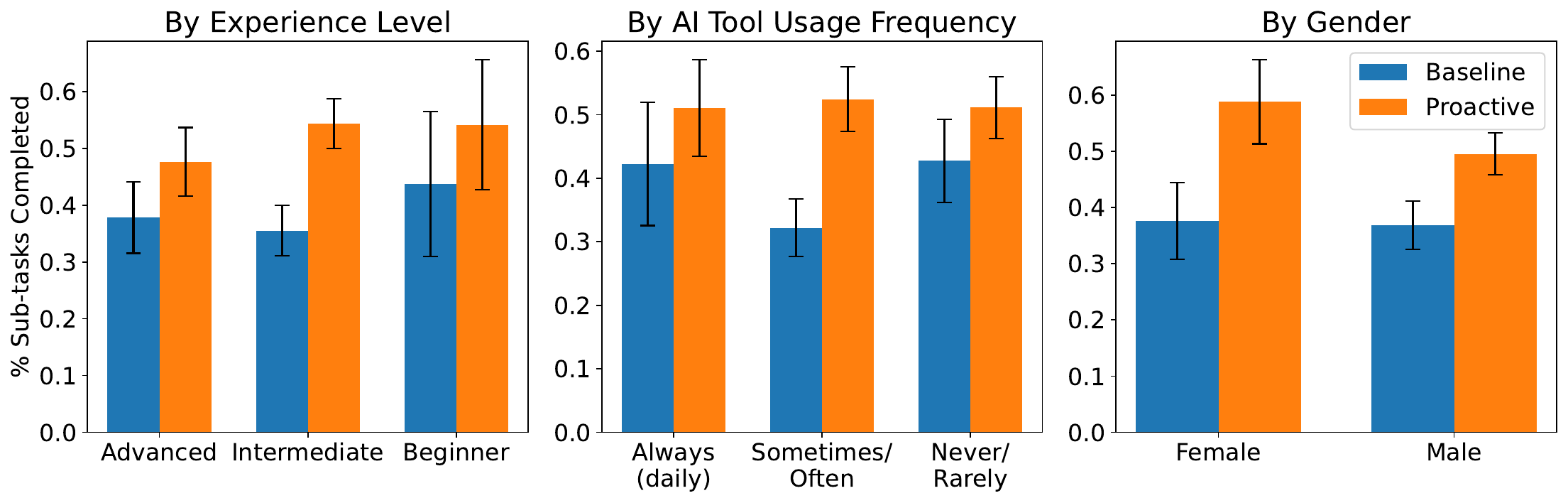}
\caption{\textbf{Percentage of sub-tasks completed correctly broken down by sub-groups.} Comparing baseline chat to proactive assistants across different sub-groups as defined by Python programming experience, AI tool usage frequency, and gender. We report mean performance (by percentage sub-tasks completed) as well as standard error).}
\Description{\textbf{Percentage of sub-tasks completed correctly broken down by sub-groups.} Comparing baseline chat to proactive assistants across different sub-groups as defined by Python programming experience, AI tool usage frequency, and gender. We report mean performance (by percentage sub-tasks completed as well as standard error).}
\label{fig:demographic}
\end{figure*}

\paragraph{Do the benefits of proactivity vary by user background?} As discussed in Section~\ref{subsec:RQ1}, we generally observe benefits across all proactive conditions.
A natural follow-up is to analyze whether user benefits vary depending on user background.
As part of background and demographic information, we collected participants' self-reported Python proficiency, self-reported AI tool experience, and gender.
For the purposes of this analysis, we combine all proactive conditions given the sample size per condition of each subgroup.
We compute the difference in performance per participant between the proactive and baseline condition and build a linear model with the background and demographic information as fixed effects.
We do not find any statistically significant effect due to Python experience or AI tool usage---this may be due to insufficient sample size per subgroup (e.g., there are only a small number of beginner programmers). 
However, we do observe a significant difference ($p=0.04$) due to gender where women tend to benefit more from proactive suggestions by 24.6\%.
This observation aligns with prior work that showed that software design could have different impacts on different genders~\citep{burnett2016gendermag}.
In Figure~\ref{fig:demographic}, we visualize the percentage of sub-tasks completed correctly broken down by each of the sub-groups.

\paragraph{What kinds of suggestions do participants tend to accept?}
As discussed in Section~\ref{subsub:accept}, of the 69 accepted suggestions, the most commonly types of accepted suggestions were those on \textit{brainstorming new functionality} and \textit{debugging (latent errors)}, with 18 occurrences each.
The least likely accepted suggestions were \textit{explaining existing code} and \textit{pointers to documentation}, with only 1 occurrence each.
Other number of occurrences of other suggestions were 12 for \textit{completing unfinished code}, 3 for \textit{debugging (runtime errors)}, 9 for \textit{adding unit tests}, and 8 for \textit{improving efficiency and modularity}.
We also observe that all accepted suggestions contain code snippets.

\end{document}